\documentclass[]{aa}
\usepackage{graphics}
\usepackage{graphicx}
\usepackage{txfonts}

\def\aj{AJ}                   
             
\def\apj{ApJ}                 
\def\apjl{ApJ}                
\def\apjs{ApJS}               
\def\ao{Appl.~Opt.}           
\def\apss{Ap\&SS}             
\def\aap{A\&A}

\def\mnras{MNRAS}

\def\solphys{Sol.~Phys.}

\def\msol{M$_{\odot}$}
\def\ysol{Y$_{\odot}$}
\def\lsol{L$_{\odot}$}
\def\teff{T$_{eff}$}
\def\mjup{M$_{\mbox{J}}$}
\def\yg{Y$_G$}
\def\al{$\alpha_{ov}$}

\begin{document}      
%
   \title{New seismic analysis of the exoplanet-host star $\mu$ Arae.} 
%
 
        \titlerunning{New seismic analysis of $\mu$ Arae}  
 
   \author{
M.~Soriano
\and S.~Vauclair 
}
 
   \offprints{M. Soriano}

   \institute{Laboratoire d'Astrophysique de Toulouse et Tarbes - UMR 5572 - Universit\'e de Toulouse - CNRS, 14, av. E. Belin, 31400 Toulouse, France} 

\mail{sylvie.vauclair@ast.obs-mip.fr}
   
\date{Received \rule{2.0cm}{0.01cm} ; accepted \rule{2.0cm}{0.01cm} }

\authorrunning{ M.~Soriano and S.~Vauclair}

\abstract
 {}
{We present detailed modelling of the exoplanet-host star $\mu$ Arae, using a new method for the asteroseismic analysis, and taking into account the new value recently derived for the Hipparcos parallax. The aim is to obtain precise parameters for this star and its internal structure, including constraints on the core overshooting.}
{We computed new stellar models in a wider range than Bazot et al. (2005), with various chemical compositions ([Fe/H] and Y), with or without overshooting at the edge of the core. We computed their adiabatic oscillation frequencies and compared them to the seismic observations. For each set of chemical parameters, we kept the model which represented the best fit to the echelle diagram. Then, by comparing the effective temperatures, gravities and luminosities of these models with the spectroscopic error boxes, we were able to derive precise parameters for this star.}
{First we find that all the models which correctly fit the echelle diagram have the same mass and radius, with an uncertainty of the order of one percent. Second, the final comparison with spectroscopic observations leads to the conclusion that besides its high metallicity, $\mu$ Arae has a high helium abundance of the order of Y~=~0.3. Knowing this allows finding precise values for all the other parameters, mass, radius and age. }
{}

\keywords{}

\maketitle
                                                                                                                                         
\section{Introduction} 

Asteroseismology of solar type stars is a powerful tool which can lead to the precise determination of the stellar parameters when associated with spectroscopic observations. Special tools have been developped for this purpose, which help constraining the internal structure of the stars from the observable acoustic frequencies. In previous papers (e.g. Vauclair et al. \cite{vauclair08}), we developped a systematic way of comparing models with observations. First, for each set of chemical composition [Fe/H] and Y, we compute evolutionary tracks and for each mass the model which best fits the observed large separation is derived. Then we keep only the model that also best fits the observed echelle diagram. We finally have a set of models correctly fitting the seismic data, with different abundances. Interestingly enough, all these models have the same mass and radius, and of course the same log g (see Vauclair et al. \cite{vauclair08} for the star $\iota$~Hor). All these models are placed in a log~$g$~-~log~\teff~diagram, together with the observed spectroscopic boxes. Finally we only keep the models which satisfy all the constraints, including seismology and spectroscopy. In this way, precise values of the stellar parameters are found.

Here we test this method on the exoplanet-host star $\mu$ Arae (HD~160691). This G3 IV-V type star is at the centre of a four-planets system: 
\begin{itemize}
  \item $\mu$ Arae b, a 1.7~\mjup~planet with an eccentric orbit (e = 0.31) and a period P~$\simeq$~638 days (Butler et al. \cite{butler01}; Jones et al. \cite{jones02}).
  \item $\mu$ Arae c, a 14 M$_{\oplus}$ planet with P~$\simeq$~9.5 days. (Santos et al. \cite{santos04b}).
  \item $\mu$ Arae d, recently discovered by Pepe et al. (\cite{pepe07}), with a period P~$\simeq$~310 days and a mass of 0.52~\mjup.
  \item $\mu$ Arae e, a long period planet (P~$\simeq$~11.3 years) with a mass of 1.81~\mjup~(Jones et al. \cite{jones02}; McCarthy et al. \cite{mccarthy04}; Pepe et al. \cite{pepe07}).
\end{itemize}

The star's overmetallicity has been established by many groups of observers: Bensby et al.~(\cite{bensby03}), Laws et al.~(\cite{laws03}), Santos et al.~(\cite{santos04a}), Santos et al.~(\cite{santos04b}) and Fischer \& Valenti~(\cite{fischer05}) (see Table~\ref{tab1}).

$\mu$ Arae was observed with the HARPS spectrometer at La Silla Observatory during eight nights in June 2004 to obtain radial velocity time series. The analysis of these data led to the discovery of up to 43 frequencies that could be identified with p-modes of degrees $\ell=0$ to 3 (Bouchy et al. \cite{bouchy05}). A detailed modelling was given by Bazot et al. (\cite{bazot05}). They computed models with two different assumptions that could explain the observed overmetallicity: overabundance of metals in the original interstellar cloud or accretion of planetary material onto the star. They tried to obtain evidence of the origin of $\mu$ Arae's overmetallicity. The results were not conclusive in that respect, as the differences were not large enough to decide between the two scenarii. Later on, other evidences were obtained that the observed overmetallicity in exoplanet host stars must be original, not due to accretion (see Castro et al. \cite{castro09}).

We computed new models, testing various values of the original metallicity and helium abundance. We compared the parameters of these models with those obtained from spectroscopy, and introduced the new value of the Hipparcos parallax, as given by van Leeuwen (\cite{leeuwen07}). We also analysed models with overshooting at the edge of the stellar core. In some of these models, negative small separations appear so that there is a crossing point in the echelle diagram for the lines $\ell=0$~-~$\ell=2$. This effect was specially discussed in Soriano \& Vauclair (\cite{soriano08}), who showed that all solar type stars go through a stage where the small separations become negative in the observable range of frequencies. Here, we present a direct application of this theoretical effect, which is used to constrain core overshooting. 

\begin{table}
\caption{Effective temperatures, gravities, and metal abundances observed for $\mu$ Arae. [Fe/H] ratios are given in dex.}
\label{tab1}
\begin{center}
\begin{tabular}{cccc} \hline
\hline
T$_{\mbox{eff}} (K)$ & log g         & [Fe/H]        & Reference \cr
 \hline
5800$\pm$100        & 4.30$\pm$0.10 & 0.32$\pm$0.10 & Bensby et al. \cite{bensby03}\cr
5811$\pm$45         & 4.42$\pm$0.06 & 0.28$\pm$0.03 & Laws et al. \cite{laws03}\cr
5798$\pm$33         & 4.31$\pm$0.08 & 0.32$\pm$0.04 & Santos et al. \cite{santos04a}\cr
5813$\pm$40         & 4.25$\pm$0.07 & 0.32$\pm$0.05 & Santos et al. \cite{santos04b}\cr
5784$\pm$44         & 4.30$\pm$0.06 & 0.29$\pm$0.03 & Fischer \& Valenti \cite{fischer05}\cr
\hline
\end{tabular}
\end{center}
\end{table}

\section{Stellar parameters}

\subsection{Spectroscopic constraints}

In Bazot et al. (\cite{bazot05}), the effective temperatures and luminosities of the models were the first parameters used for comparisons with the observations. We now prefer to compare the models with the spectroscopic observations in a more consistent way, by using the ``triplets'' \teff, log~$g$ and [Fe/H]. These three parameters are indeed consistently given be the same observers, while the luminosities are obtained in a different way, using Hipparcos parallaxes. Here we compare the luminosities as a second step. For this reason, we do not use exactly the same set of references as given by Bazot et al. (\cite{bazot05}), as we only keep those which give constraints on the surface gravity (see Table~\ref{tab1}). 

\subsection{The luminosity of $\mu$ Arae}

The visual magnitude of $\mu$ Arae is $V=5.15$ (SIMBAD Astronomical data base). Van Leeuwen (\cite{leeuwen07}) carried out a new analysis of the Hipparcos data. He derived a new value of the parallax of $\mu$ Arae: $\pi = 64.48 \pm 0.31$ mas, which is lower than the first one derived by Perryman et al.~(\cite{perryman97}) and has a reduced error. Using this new value, we deduced an absolute magnitude of $M_{V}=4.20 \pm 0.04$. 

From Table~\ref{tab1}, we have an average value of effective temperature for $\mu$ Arae of \teff~=~5800~$\pm$~100~K. Using the tables of Flower (\cite{flower96}), we obtained $BC=-0.08\pm0.02$ for the bolometric correction.
With a solar absolute magnitude of $M_{bol,\odot}=4.75$ (Lejeune et al. \cite{lejeune98}), we deduced for $\mu$ Arae a luminosity of log~(L/\lsol)$=0.25\pm0.03$. This value is lower than the one derived by Bazot et al (\cite{bazot05}) in their analysis of this star, namely log~(L/\lsol)$=0.28\pm0.0012$.

\subsection{Seismic constraints}

The star $\mu$ Arae was observed with the HARPS spectrograph, dedicated to the search for exoplanets by the means of precise radial velocity measurements. HARPS is installed on the 3.6m-ESO telescope at la Silla Observatory, Chile.

These measurements led to the identification of 43 p-modes of degrees $\ell=0$ to $\ell=3$ with frequencies between 1.3 and 2.6 mHz (Bouchy et al. \cite{bouchy05}). The frequency resolution of the time-series was 1.56~$\mu$Hz, and the uncertainty on the oscillation modes has been evaluated to 0.78~$\mu$Hz. The mean large separation between two modes of consecutive order, computed in the observed range of frequencies, is $\Delta\nu_0=$~90~$\mu$Hz, with an uncertainty of 1.1 $\mu$Hz. The small separations present variations which are clearly visible in the echelle diagram. Their average value in the observed range of frequencies is $<\delta\nu>$=~5.7~$\mu$Hz.

\section{Evolutionary tracks and models}

We computed evolutionary tracks with the TGEC code (Hui Bon Hoa \cite{hui08}; Richard et al.~\cite{richard96}), with the OPAL equations of state and opacities (Rogers \& Nayfonov \cite{rogers02}; Iglesias \& Rogers \cite{iglesias96}) and the NACRE nuclear reaction rates (Angulo et al. \cite{angulo99}). For all models, the microscopic diffusion was included as in Paquette et al. (\cite{paquette86}) and Michaud et al. (\cite{michaud04}). The treatment of the convection is done in the framework of the mixing length theory and the mixing length parameter is adjusted as in the Sun: $\alpha=1.8$ (Richard et al. \cite{richard04}). We also computed cases with $\alpha=1.7$ and $\alpha=1.9$ to test the corresponding uncertainties. We found that for 1.10 \msol~and $\alpha=1.7$, the track is very close to that of 1.12 \msol~and $\alpha=1.8$. On the other side, the 1.10 \msol~and $\alpha=1.9$ track is close to the 1.08 \msol~and $\alpha=1.8$ one. This was taken into account in the determination of the uncertainties on the final results (Sect. 5).

From the literature, we found two values for the metallicity of $\mu$ Arae: [Fe/H]=0.29 (Laws et al. \cite{laws03}, Fischer \& Valenti \cite{fischer05}) and [Fe/H]=0.32 (Bensby et al. \cite{bensby03}, Santos et al. \cite{santos04a} and \cite{santos04b}). For each value of [Fe/H], we first computed two series of evolutionary tracks for two different values of the helium abundance:
\begin{itemize}
\item a helium abundance increasing with Z as given by the Isotov \& Thuan (\cite{isotov04}) law for the chemical evolution of the galaxies: we label this value \yg.
\item a solar helium abundance \ysol~=~0.2714
\end{itemize}

\begin{figure*}
\begin{center}
\includegraphics[angle=0,totalheight=4.8cm,width=7.0cm]{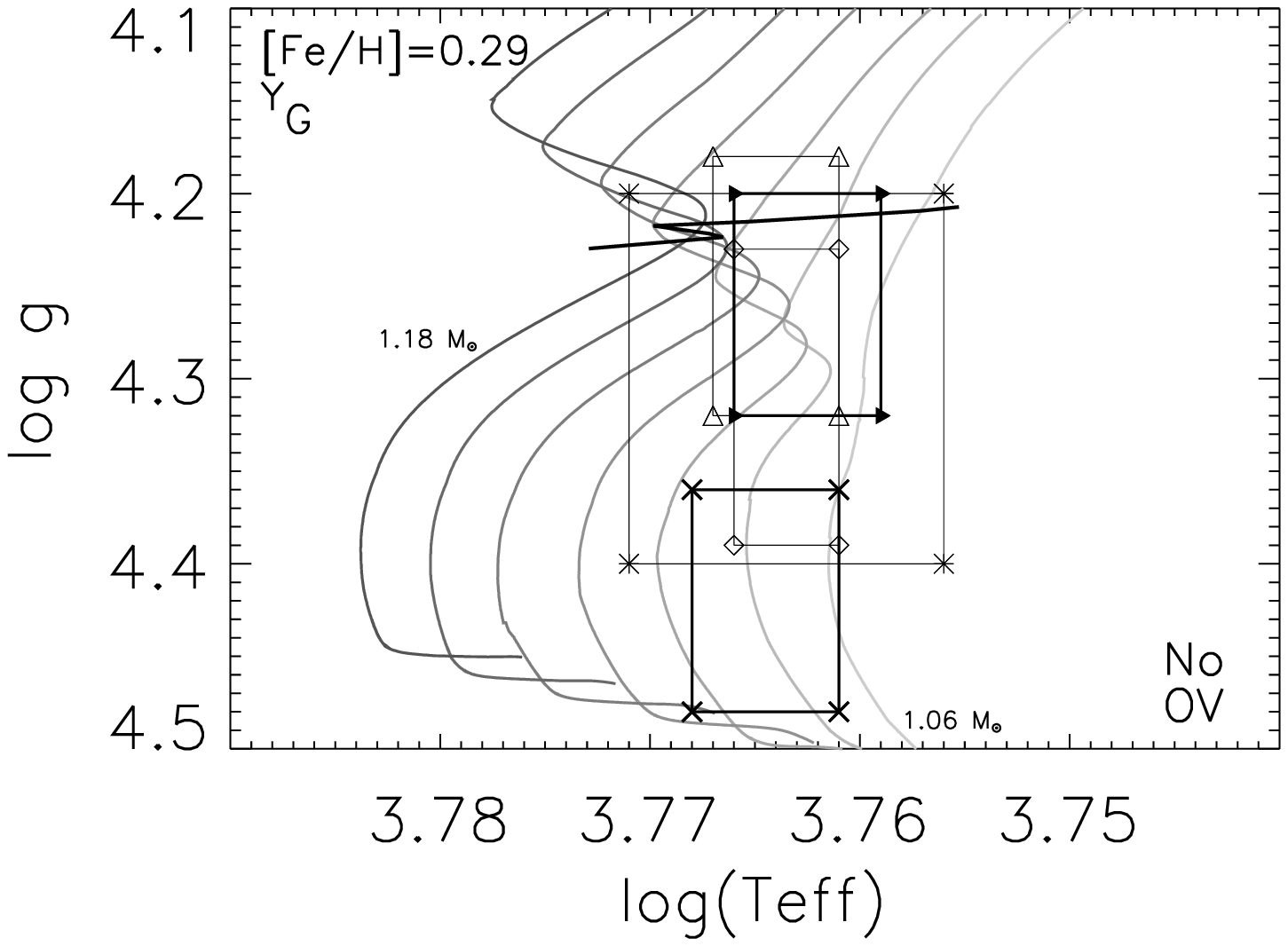}\includegraphics[angle=0,totalheight=4.8cm,width=7.0cm]{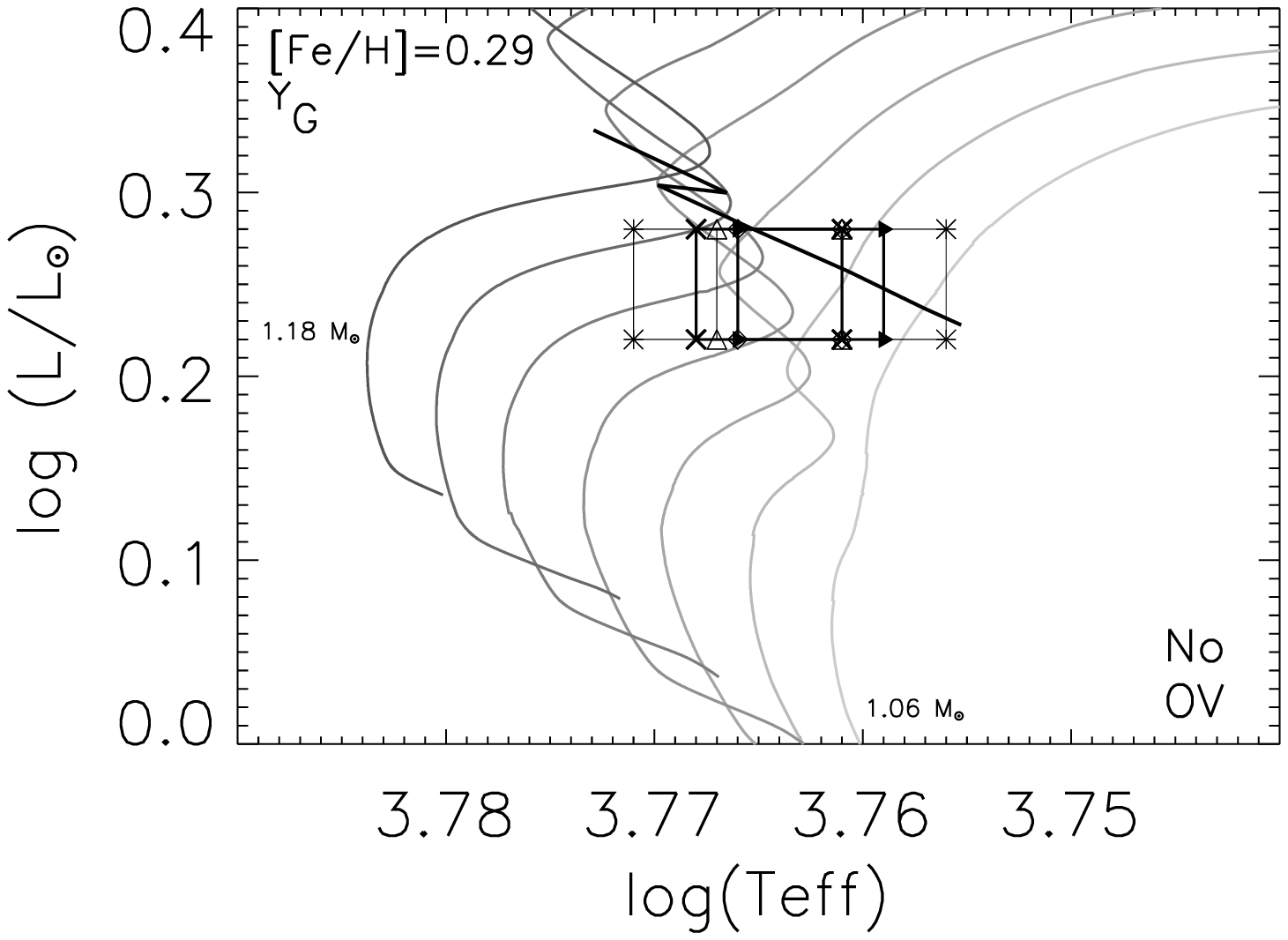}
\includegraphics[angle=0,totalheight=4.8cm,width=7.0cm]{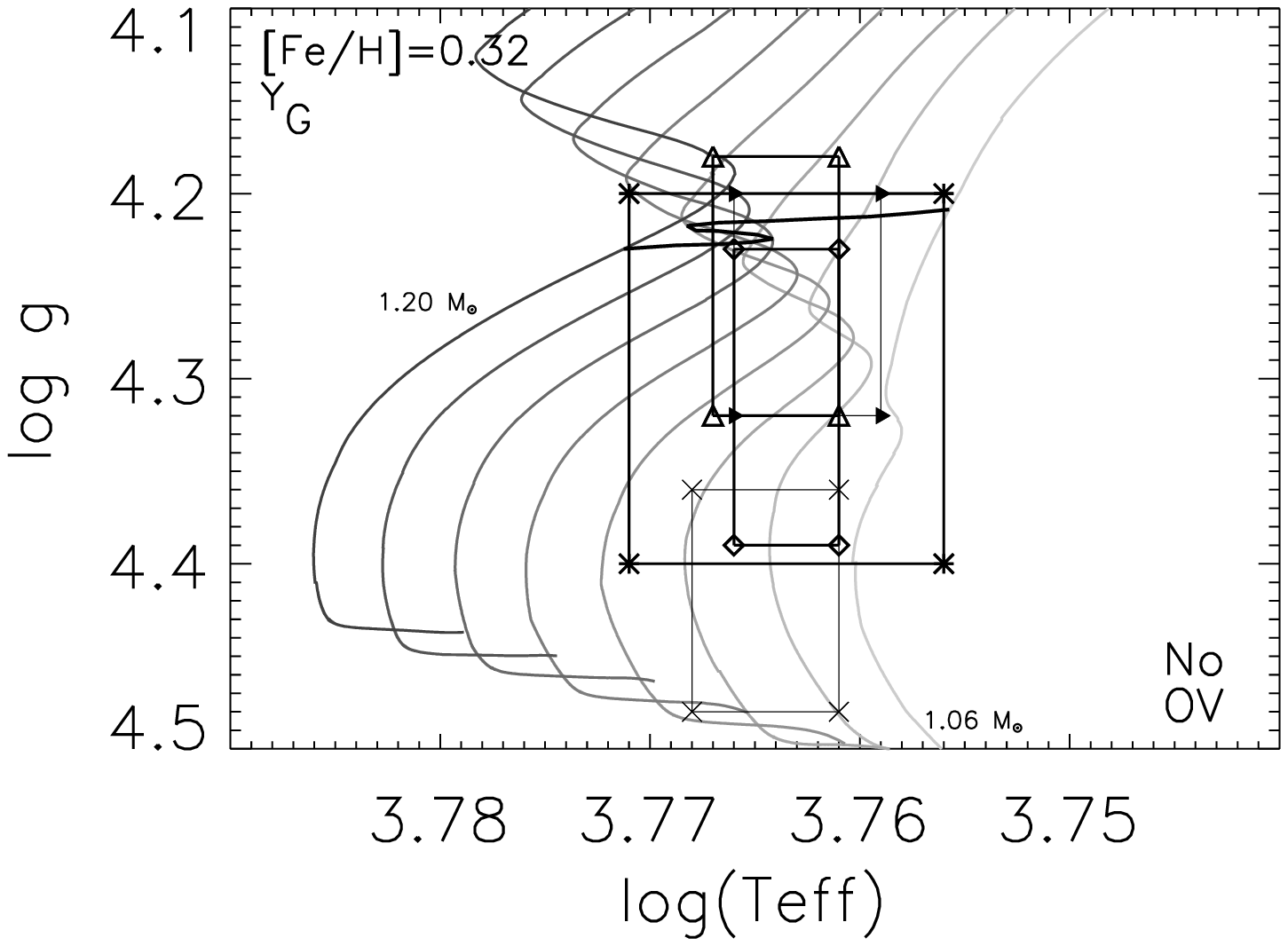}\includegraphics[angle=0,totalheight=4.8cm,width=7.0cm]{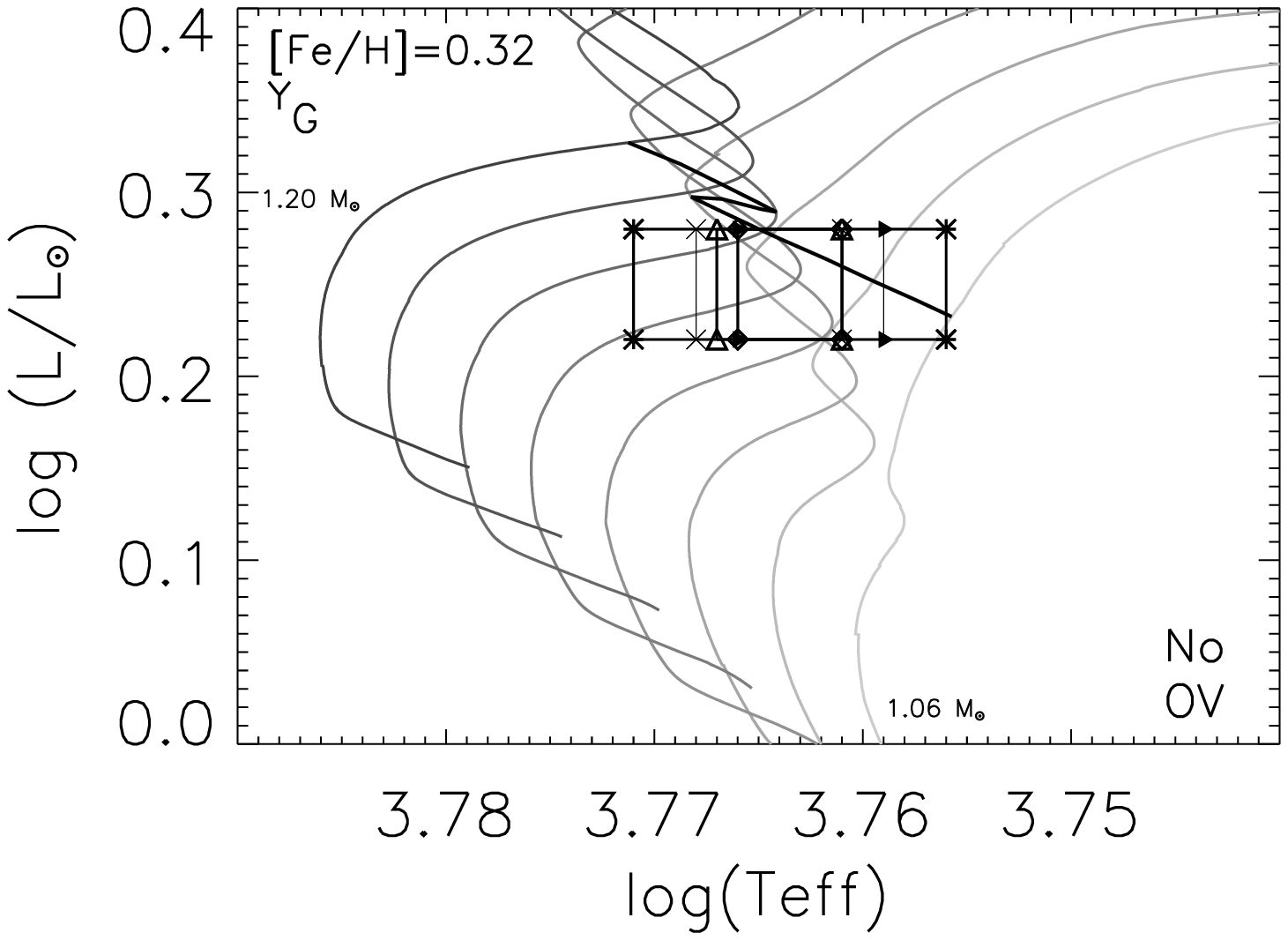}
\includegraphics[angle=0,totalheight=4.8cm,width=7.0cm]{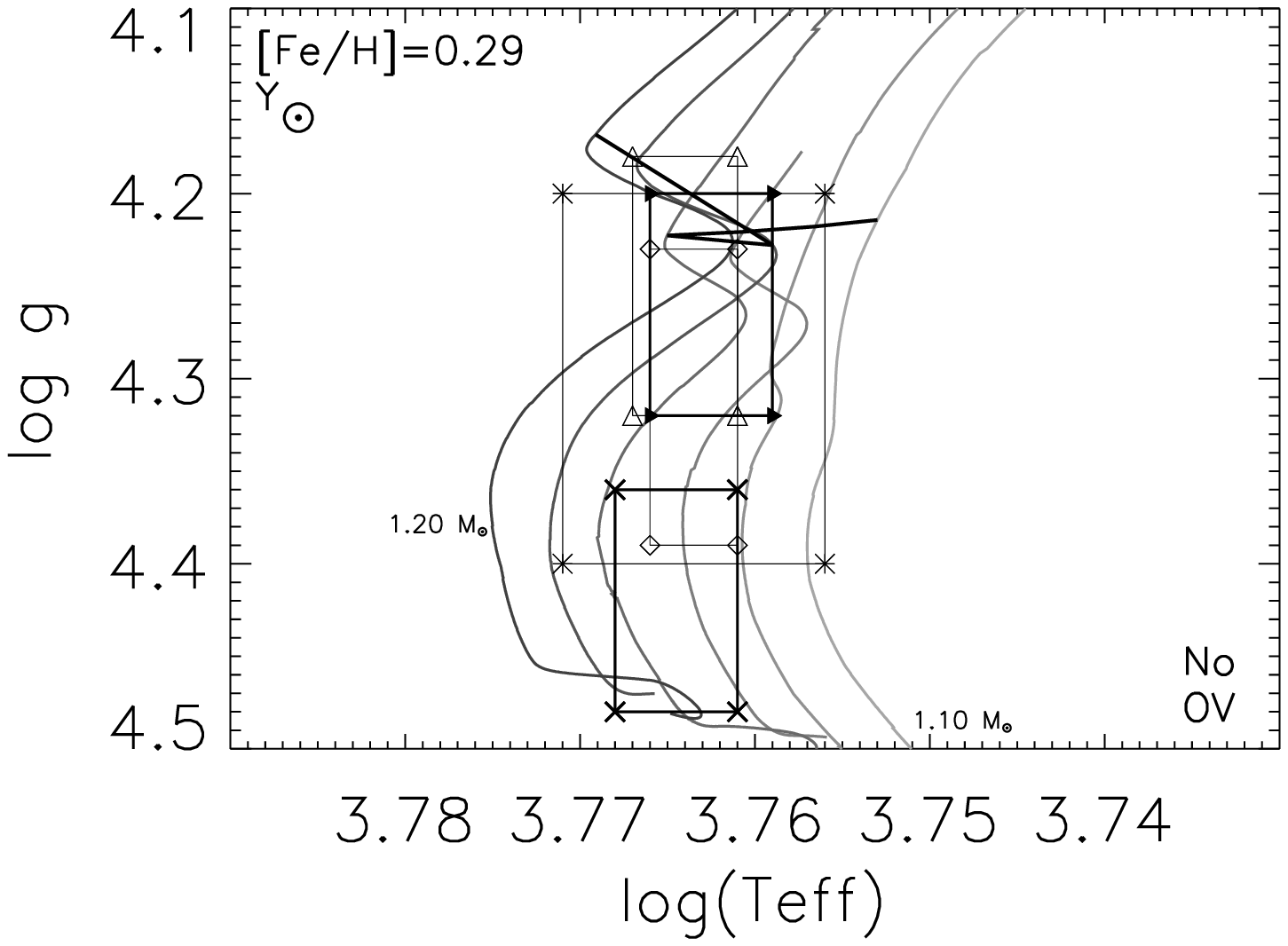}\includegraphics[angle=0,totalheight=4.8cm,width=7.0cm]{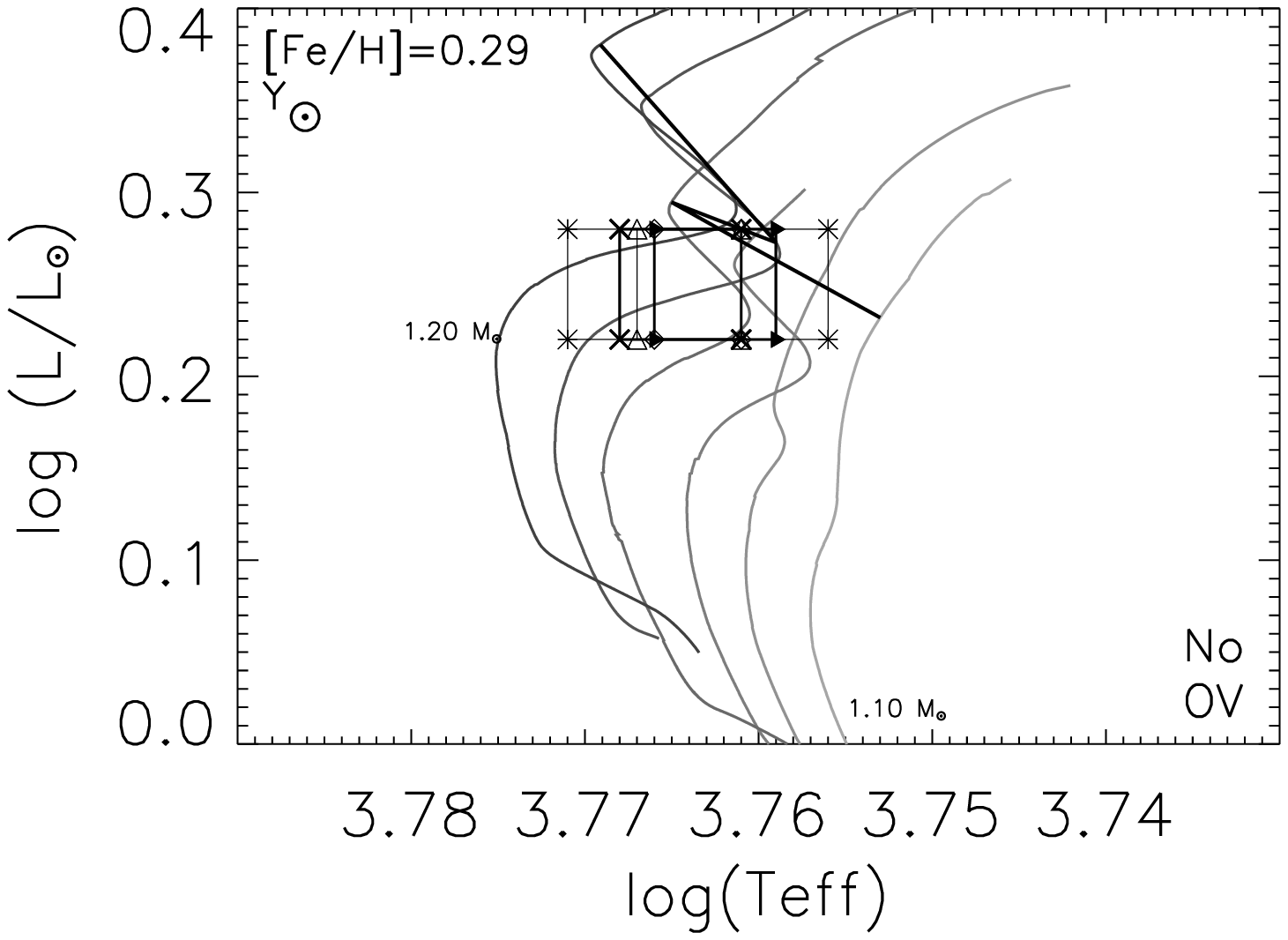}
\includegraphics[angle=0,totalheight=4.8cm,width=7.0cm]{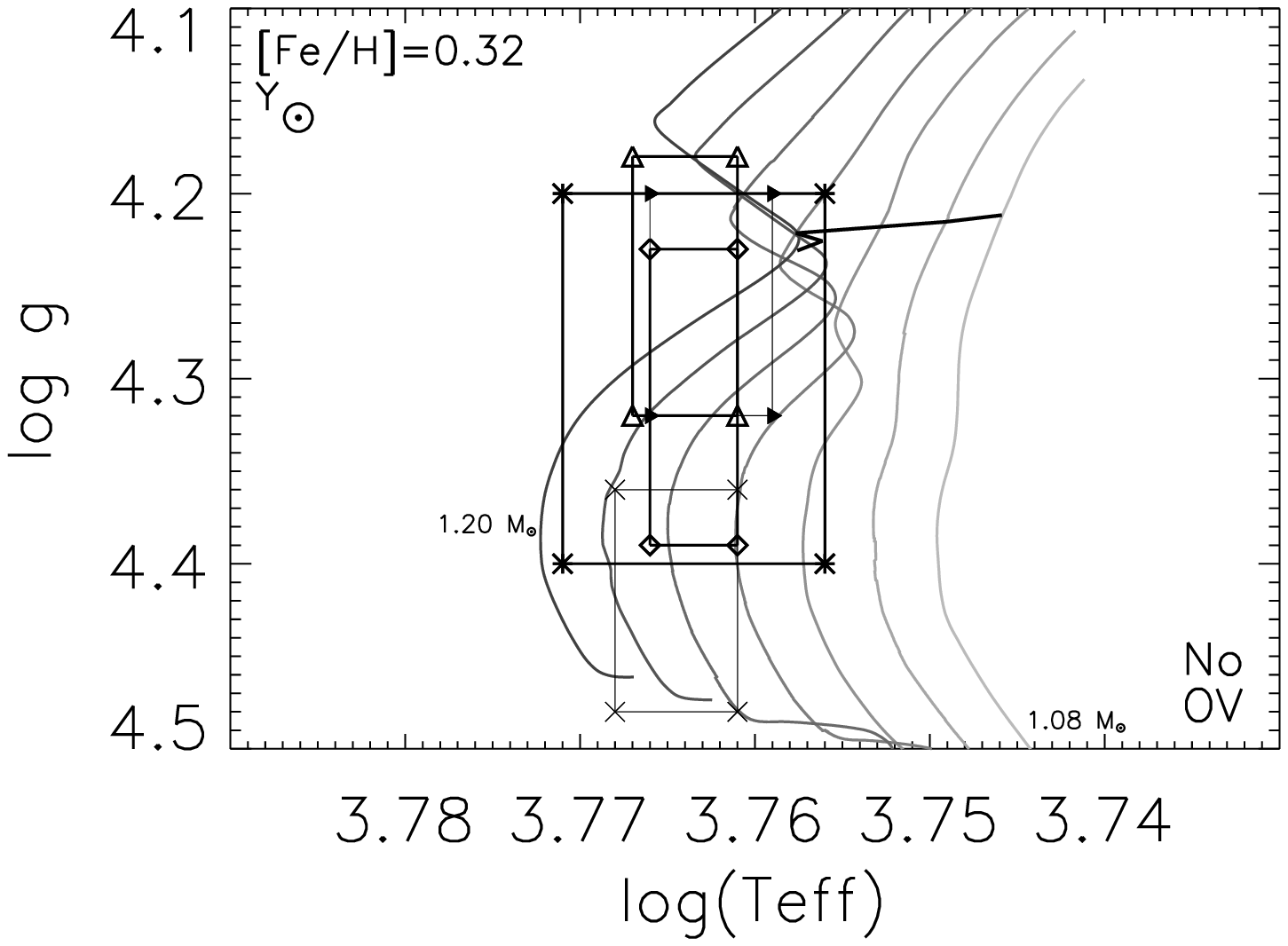}\includegraphics[angle=0,totalheight=4.8cm,width=7.0cm]{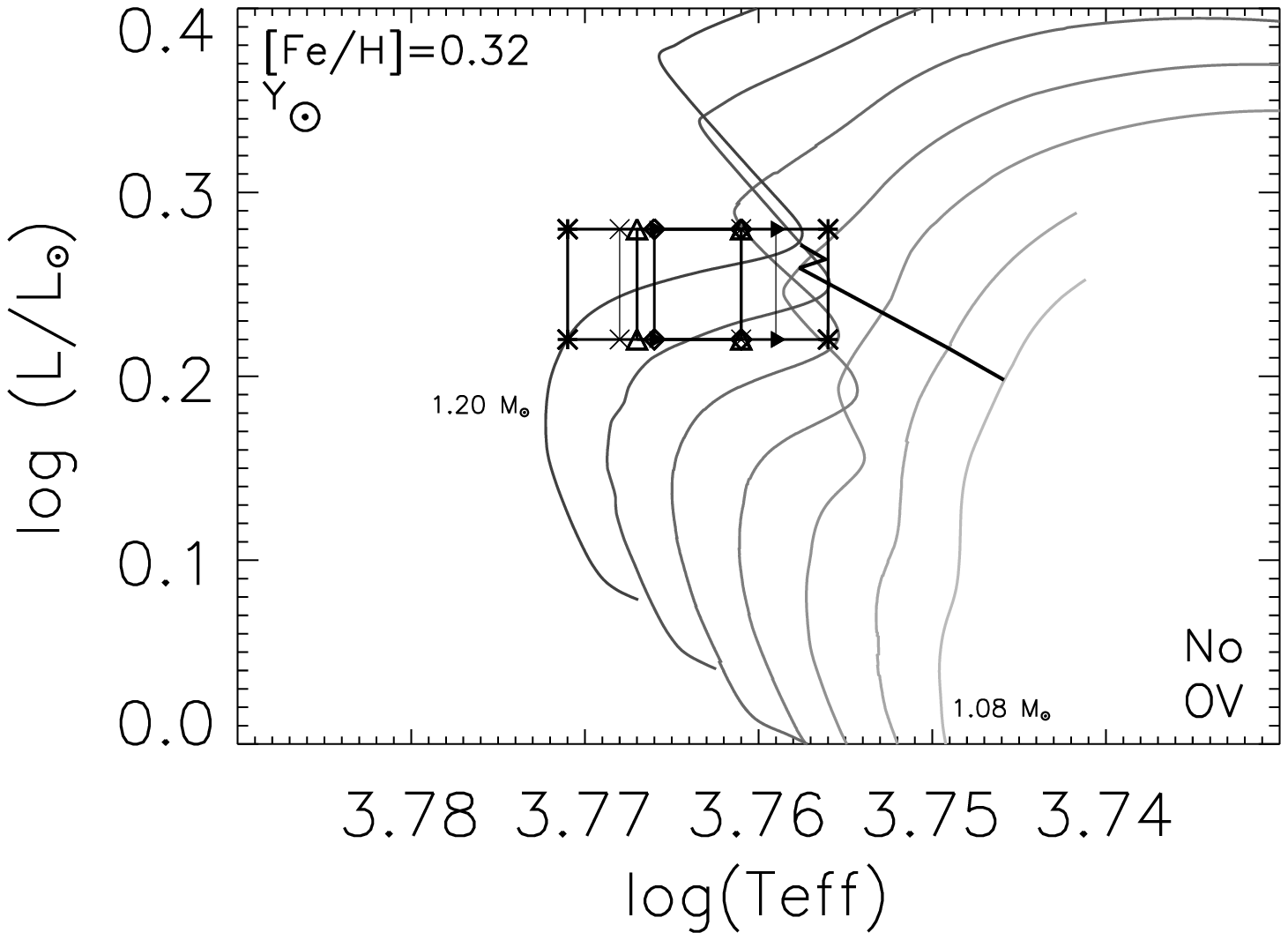}
\end{center}
\caption{Evolutionary tracks in the log~$g$~-~log~\teff~(left panels) and log~(L/\lsol)~-~log~\teff~planes (right panels), for the two values of metallicities found in the literature: [Fe/H]=0.29 and 0.32. The five error boxes shown are from Bensby et al. \cite{bensby03} (asterisks), Laws et al. \cite{laws03} (crosses), Santos et al. \cite{santos04a} (diamonds), Santos et al \cite{santos04b} (triangles), and Fischer \& Valenti \cite{fischer05} (black triangles). The represented masses are 1.06, 1.08, 1.10, 1.12, 1.14, 1.16, 1.18, and 1.20 \msol. The horizontal thick line represents the iso-$\Delta\nu_0$ 90~$\mu$Hz line, i.e. those models that have a mean large separation of 90~$\mu$Hz.}
\label{fig1}
\end{figure*}

The results obtained in the log~$g$ - log~\teff~ and log~(L/\lsol)~-~log\teff~planes are displayed in Fig.\ref{fig1}. Each graph corresponds to one value of the metallicity. On each graph, the five error boxes are plotted, but those that correspond to the same value of [Fe/H] as given by the groups of observers are drawn in thicker lines.

We then computed two similar series including overshooting at the edge of the convective core. Here overshooting is simply described as an extension of the convective core by a length \al~H$_P$, where \al~is the overshooting parameter, and H$_P$ the pressure scale height. In these two series, the overshooting parameter is fixed at \al~=~0.2 (Fig.~\ref{fig6}). Finally, we tested more precisely the seismic constraints on overshooting by varying \al~ in small steps between 0.0 and 0.2 for models of masses 1.1~\msol~(Fig.~\ref{fig7}).

\section{Seismic tests and results}

\subsection{Models without overshooting}

We computed the adiabatic oscillation frequencies for a large number of models along each evolutionary track with the PULSE code (Brassard \& Charpinet \cite{brassard08}). These frequencies were computed for degrees $\ell=0$ to $\ell=3$, and for radial orders typically ranging from 4 to 100. For each track, we selected the model that has a mean large separation of 90~$\mu$Hz, computed in the same frequency range as the observed one. We plotted on each graph (Fig.~\ref{fig1}) the corresponding iso-$\Delta\nu_0$~90~$\mu$Hz line. The parameters of these models are given in Tables~\ref{tab2} to \ref{tab5}. We can check that all the models that fit the same large separation present the same value of the parameter $g/R$ or $M/R^3$ where $g$, M and R are respectively the gravity, mass and radius of the star.

This is because the large separation approximately varies like $c/R$, where c represents the mean sound velocity in the star, which itself approximately varies like $\sqrt{T/\mu}$. With the usual scaling for the stellar temperature $T\simeq\mu M/R$, we obtain that $(c/R)^2$ varies like $M/R^3$, independently of the chemical composition.

Note on Fig.~\ref{fig1} that the iso-$\Delta\nu_0$~90~$\mu$Hz lines never cross Laws et al.~(\cite{laws03}) or Santos et al.~(\cite{santos04a}) error boxes. Models with $\Delta\nu$~=~90~$\mu$Hz have a value of log~$g$ much lower than the one derived by these two groups of observers.

Kjeldsen et al (\cite{kjeldsen08}) discussed possible corrections on the frequencies due to near-surface effects. They gave a parameterisation formula including three parameters, in which two of them can be deduced from the third one. As this parameter is unknown, they suggest using the same value for all solar type stars as deduced for the Sun in a first approximation. They find that the induced shifts increase with the frequency values, so that it can slightly modify the computed mean large separation, according to the frequency range. In our echelle diagrams for $\mu$ Arae, the models and the observations correctly fit when we do detailed mode to mode comparisons (Fig.~\ref{fig4}). We conclude that these near-surface effects are negligible in the observed range of frequencies, i.e. at least up to 2.3 mHz. 

For each value of [Fe/H] and Y we searched, among the models which fit a mean large separation of 90~$\mu$Hz, the one which also gives the best fit to the observational echelle diagram. In order to find this ``best model'', we performed a $\chi^2$ minimization for the echelle diagram of the models. Tables~\ref{tab2} to \ref{tab5} show the parameters of some models close to the minimum of $\chi^2$. We used the expression
\begin{eqnarray}
\chi^2=\frac{1}{N} \sum_{i=1}^N \left ( \frac{\nu_{th}(i)-\nu_{obs}(i)}{\sigma_{obs}} \right )^2,
\end{eqnarray}
where $\nu_{obs}(i)$ is the observational modulated frequency, $\nu_{th}(i)$ its theoretical counterpart, and $\sigma_{obs}$ the uncertainty of the observed frequencies, for which we used the value given by Bouchy et al. (\cite{bouchy05}): 0.78~$\mu$Hz.

\begin{figure*}
\begin{center}
\includegraphics[angle=0,totalheight=6.8cm,width=10cm]{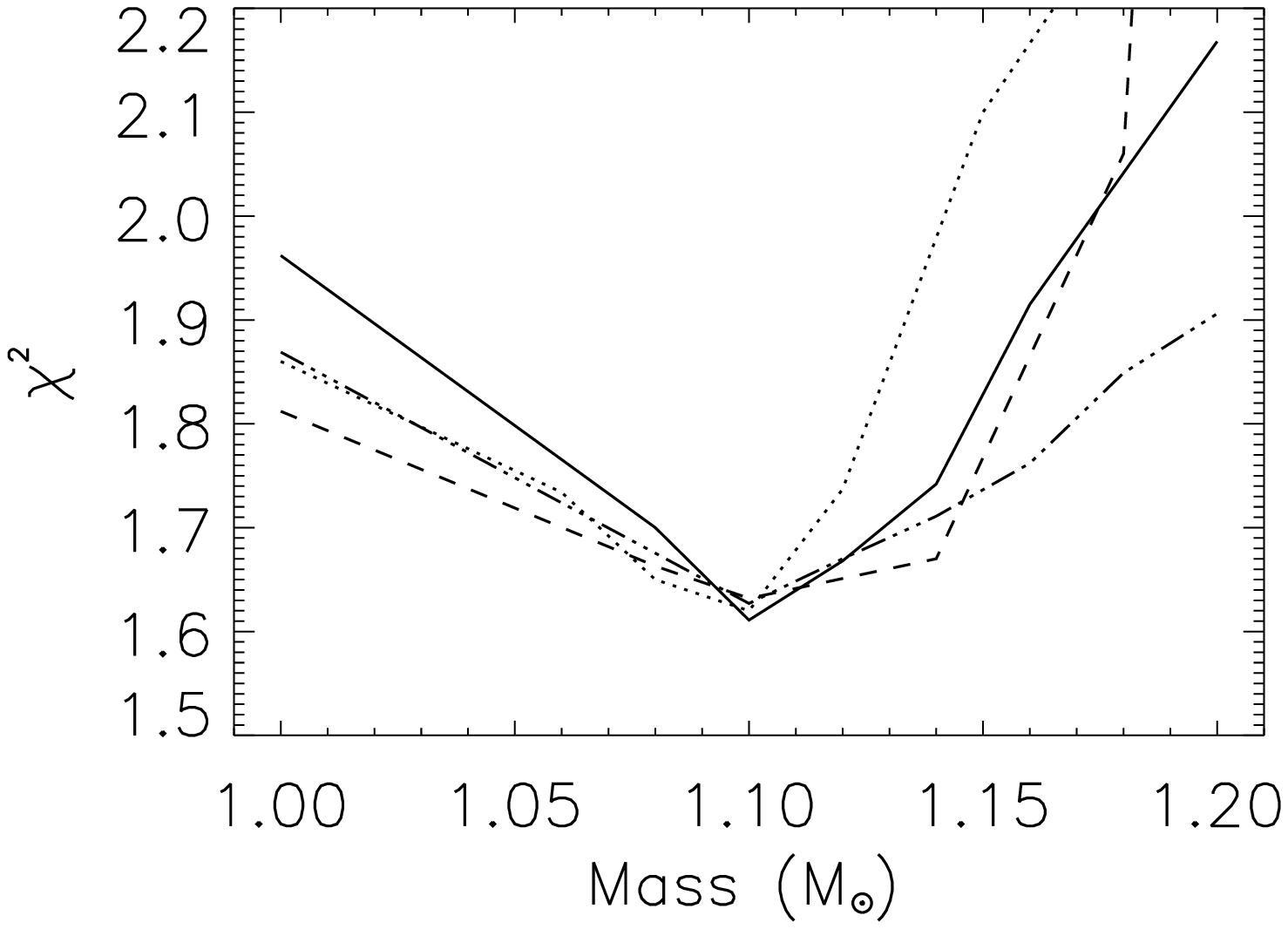}
\caption{Evolution of the $\chi^2$ value with the mass of the models, between 1~\msol~and 1.20~\msol, for the four sets of chemical composition: [Fe/H]=0.32, Y$_G$ (solid line), [Fe/H]=0.29, Y$_G$ (dotted line), [Fe/H]=0.32, \ysol~(dashed line), and [Fe/H]=0.29, \ysol~(dotted-dashed line).}
\label{fig2}
\end{center}
\end{figure*}

\begin{figure*}
\begin{center}
\includegraphics[angle=0,totalheight=5.5cm,width=8cm]{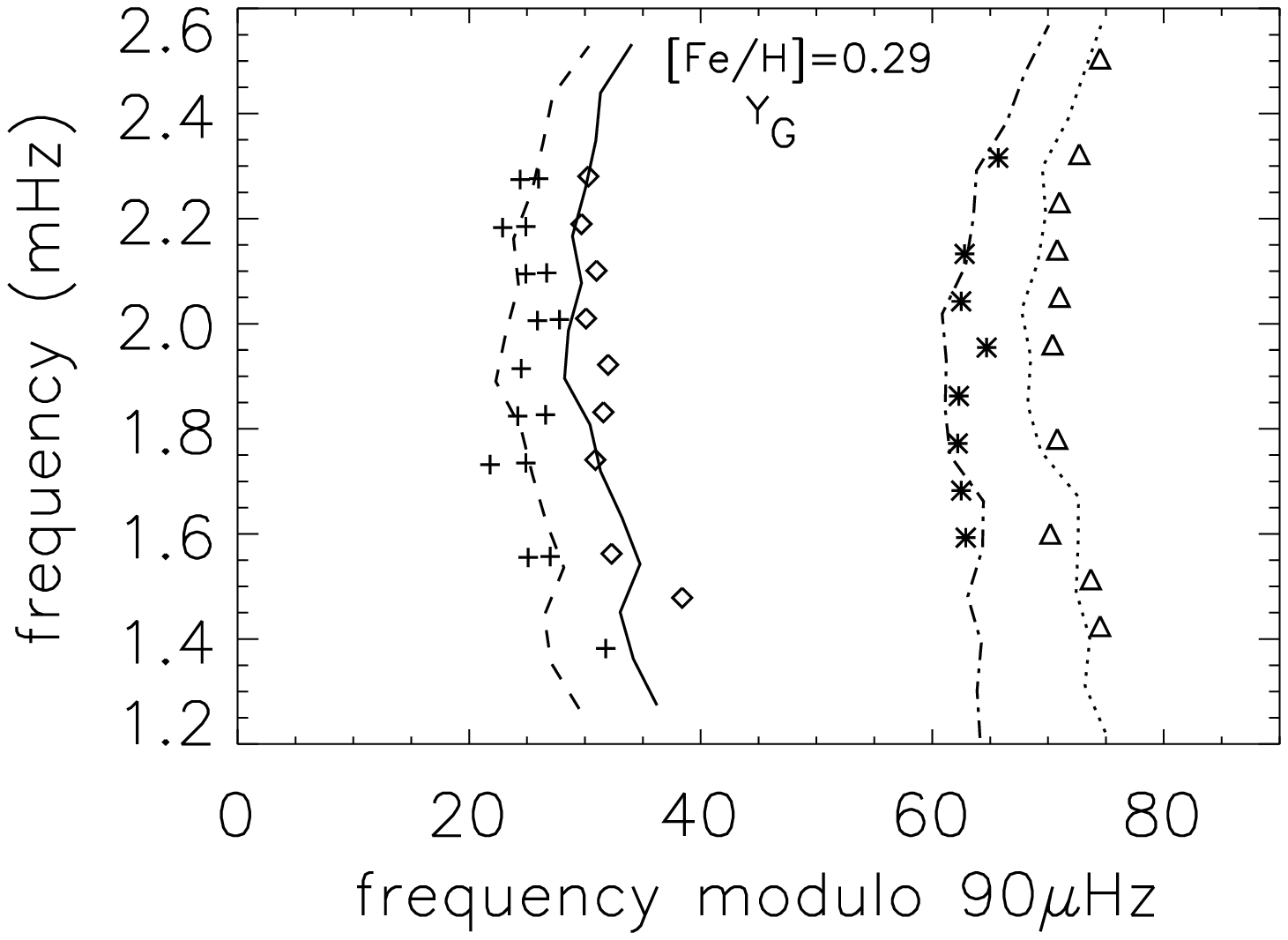}\includegraphics[angle=0,totalheight=5.5cm,width=8cm]{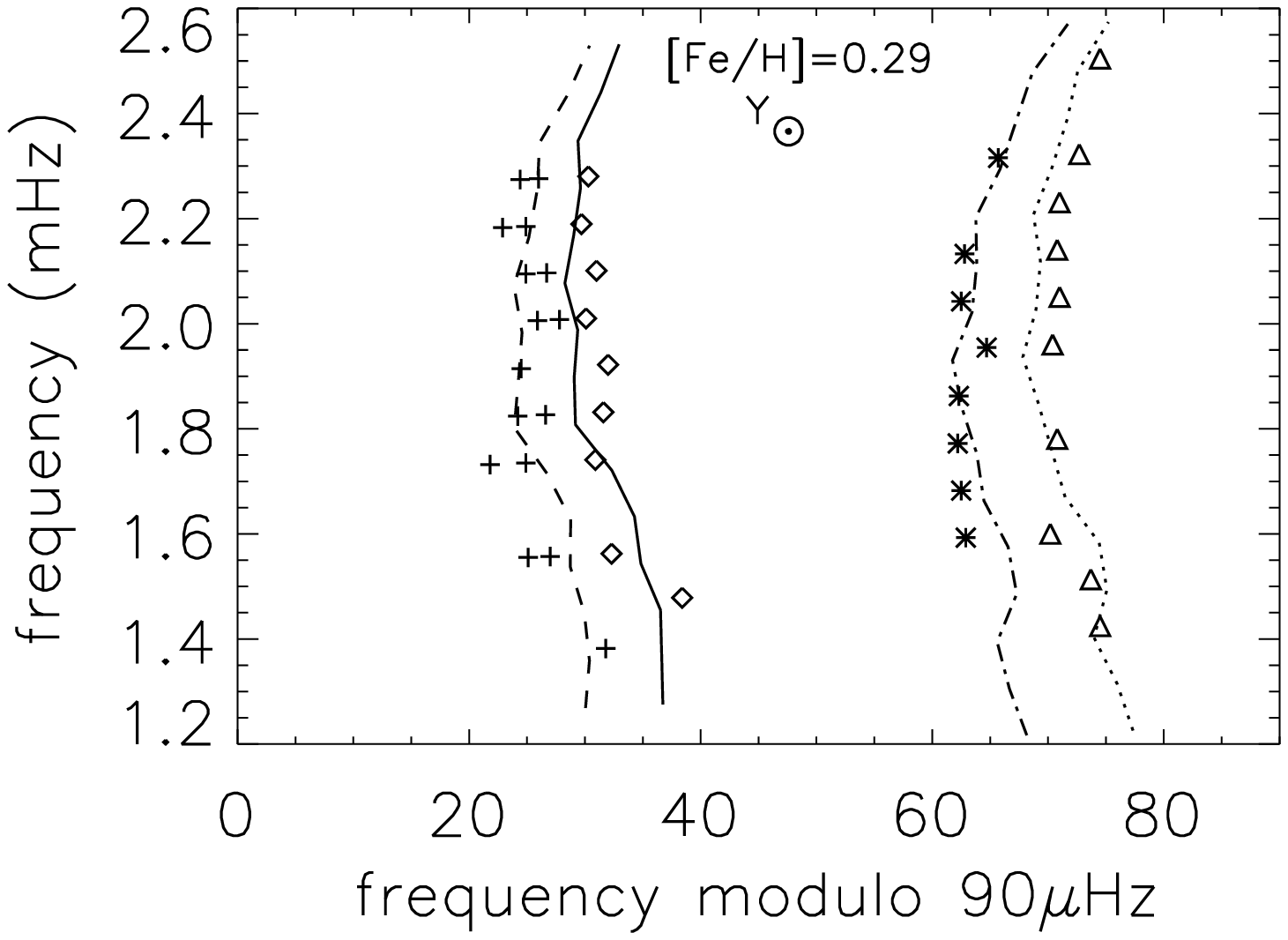}
\includegraphics[angle=0,totalheight=5.5cm,width=8cm]{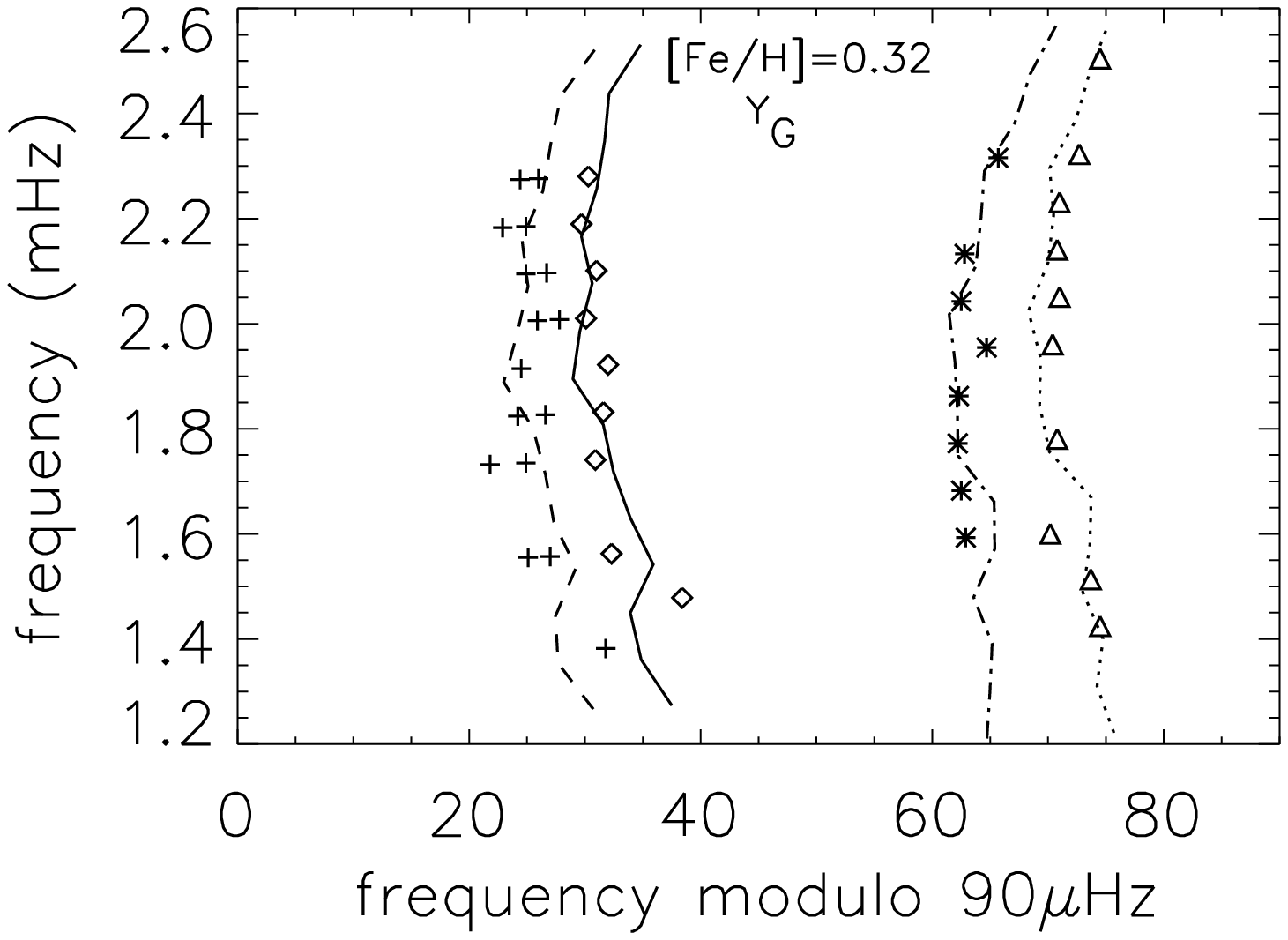}\includegraphics[angle=0,totalheight=5.5cm,width=8cm]{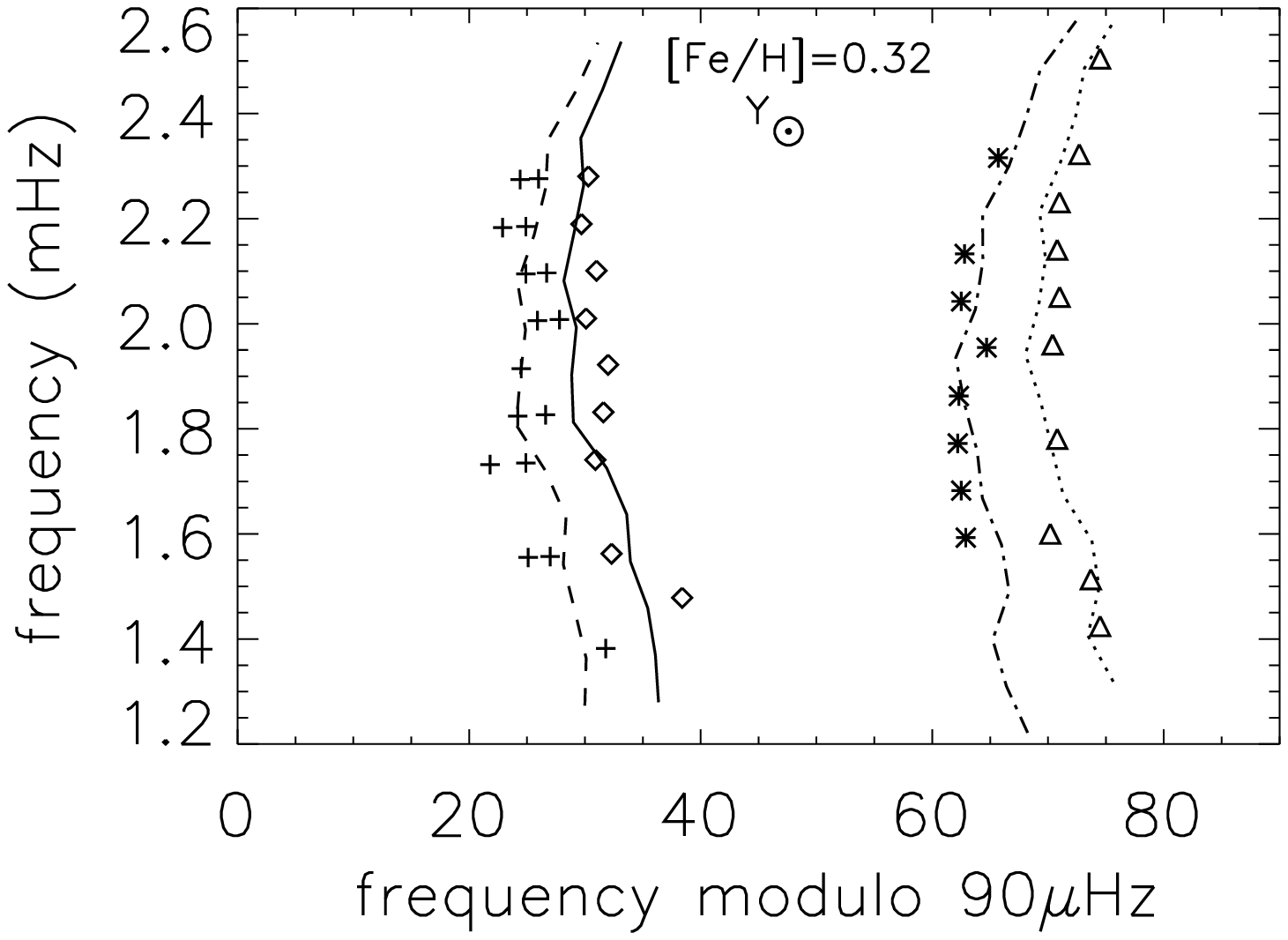}
\caption{Echelle diagrams for the best models with a metallicity of [Fe/H]=0.29 (upper panels) and [Fe/H]=0.32 (lower panels), and an helium abundance \yg (left panels) and Y$_\odot$ (right panels). The lines represent the theoretical frequencies and the symbols their observational counterpart. Solid lines and diamonds are for $\ell$=0, dotted lines and triangles are for $\ell$=1, dashed lines and crosses are for $\ell$=2, and dotted-dashed lines and asterisks are for $\ell$=3.}
\label{fig3}
\end{center}
\end{figure*}

\begin{figure*}[btp!]
\begin{center}
\includegraphics[angle=0,totalheight=6.8cm,width=10cm]{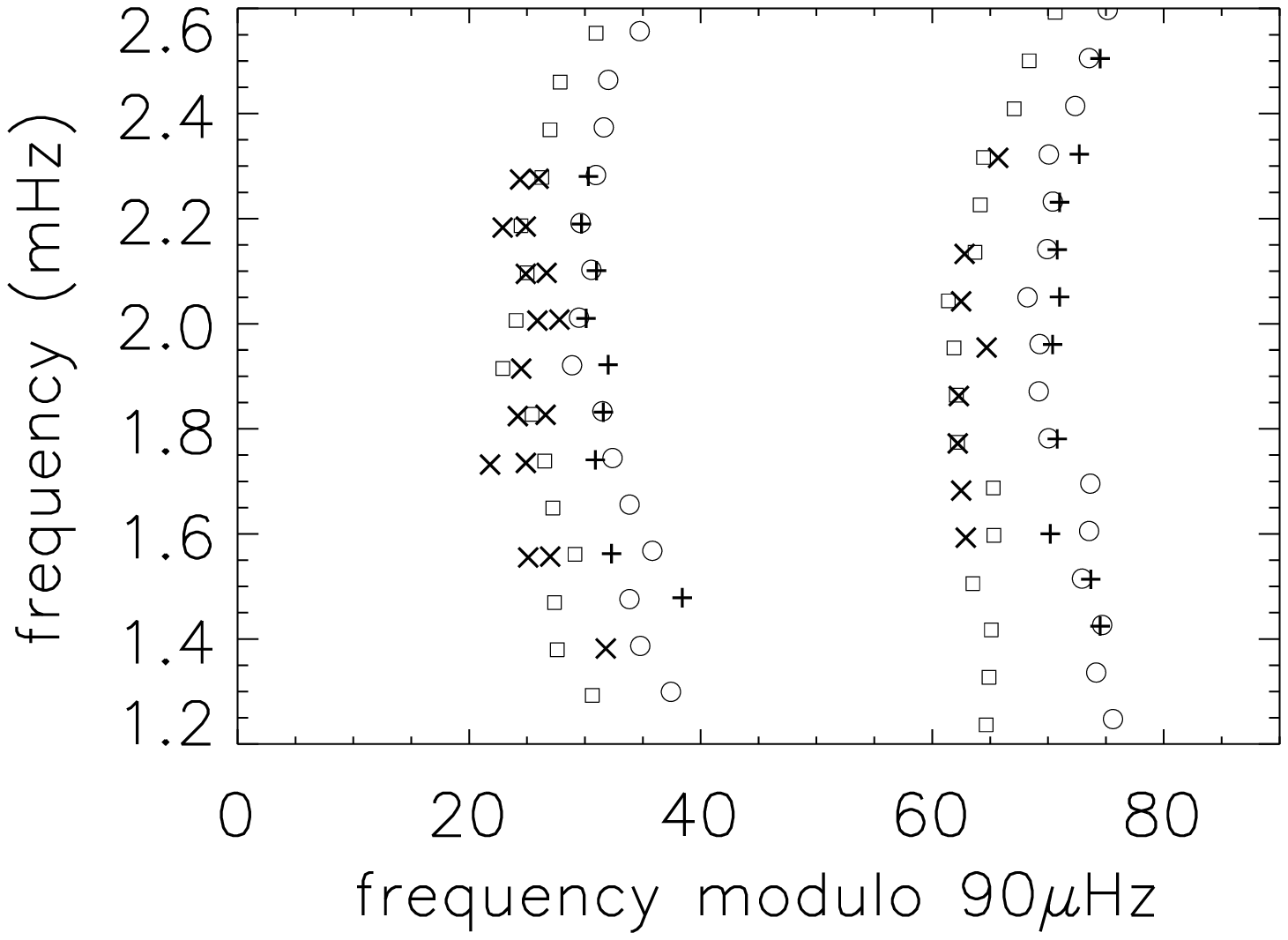}
\caption{Echelle diagram for the best model with a metallicity of [Fe/H]=0.32 and a helium abundance Y$_G$. Crosses and plus signs are for the observed frequencies, open circles and squares are for their theoretical counterpart.}
\label{fig4}
\end{center}
\end{figure*}

For all sets of chemical composition, the $\chi^2$ values present a minimum for models of mass M~=~1.10~\msol, radius R~=~9.46x10$^{10}$~cm, and gravity log~$g$~=~4.21 (Fig.~\ref{fig2}). These models represent the four ``best models'', one for each couple ([Fe/H], Y). Their echelle diagrams are displayed in Fig.~\ref{fig3}. For a better visibility of the figures, we give the observed mode frequencies separately and the modelled ones as lines. It is remarkable that their masses, radii and thus gravities are identical and has important consequences, which will be discussed below.

\begin{table*}
\caption{Characteristics of some overmetallic models with [Fe/H]=0.32 and Y$_G$}
\label{tab2}
\begin{flushleft}
\begin{tabular}{ccccccccc} \hline
\hline
Mass &  Age (Gyr) & log $g$ & log T$_{\mbox{eff}}$ & log L/\lsol & R (x10$^{10}$ cm) & M/R$^3$ & $\chi^2$\cr
\hline \hline
1.08 &  7.112     & 4.2121  & 3.7594               & 0.2519     & 9.40               & 2.58    & 1.700\cr
1.10 &  6.318     & 4.2149  & 3.7644               & 0.2770     & 9.46               & 2.58    & 1.611\cr
1.12 &  5.748     & 4.2172  & 3.7682               & 0.2975     & 9.52               & 2.58    & 1.668\cr
1.14 &  5.387     & 4.2200  & 3.7667               & 0.2967     & 9.57               & 2.58    & 1.742\cr
1.16 &  4.953     & 4.2246  & 3.7642               & 0.2894     & 9.61               & 2.59    & 1.915\cr
\hline
\end{tabular}
\end{flushleft}
\end{table*}

\begin{table*}
\caption{Characteristics of some overmetallic models with [Fe/H]=0.29 and Y$_G$}
\label{tab3}
\begin{flushleft}
\begin{tabular}{ccccccccc } \hline
\hline
Mass &  Age (Gyr) & log $g$ & log T$_{\mbox{eff}}$ & log L/\lsol & R (x10$^{10}$ cm) & M/R$^3$ & $\chi^2$\cr
\hline \hline
1.06 &  7.916     & 4.2064  & 3.7571               & 0.2371     & 9.35               & 2.58    & 1.734\cr
1.08 &  7.152     & 4.2092  & 3.7607               & 0.2569     & 9.41               & 2.58    & 1.650\cr
1.10 &  6.367     & 4.2152  & 3.7653               & 0.2801     & 9.46               & 2.58    & 1.620\cr
1.12 &  5.716     & 4.2145  & 3.7698               & 0.3039     & 9.52               & 2.58    & 1.737\cr
1.15 &  5.177     & 4.2215  & 3.7672               & 0.3006     & 9.60               & 2.58    & 2.100\cr
\hline
\end{tabular}
\end{flushleft}
\end{table*}

\begin{table*}
\caption{Characteristics of some overmetallic models with [Fe/H]=0.32 and \ysol}
\label{tab4}
\begin{flushleft}
\begin{tabular}{ccccccccc} \hline
\hline
Mass &  Age (Gyr) & log $g$ & log T$_{\mbox{eff}}$ & log L/\lsol & R (x10$^{10}$ cm) & M/R$^3$ & $\chi^2$\cr
\hline \hline
1.08 &  9.561     & 4.2088  & 3.7459               & 0.1981     & 9.41               & 2.58    & 1.663\cr
1.10 &  8.678     & 4.2120  & 3.7491               & 0.2152     & 9.46               & 2.58    & 1.632\cr
1.14 &  7.002     & 4.2183  & 3.7576               & 0.2589     & 9.56               & 2.59    & 1.670\cr
1.18 &  6.073     & 4.2230  & 3.7562               & 0.2636     & 9.68               & 2.58    & 2.060\cr
1.20 &  5.512     & 4.2281  & 3.7576               & 0.2714     & 9.70               & 2.60    & 3.572\cr
\hline
\end{tabular}
\end{flushleft}
\end{table*}

\begin{table*}
\caption{Characteristics of some overmetallic models with [Fe/H]=0.29 and \ysol}
\label{tab5}
\begin{flushleft}
\begin{tabular}{ccccccccc} \hline
\hline
Mass &  Age (Gyr) & log $g$ & log T$_{\mbox{eff}}$ & log L/\lsol & R (x10$^{10}$ cm) & M/R$^3$ & $\chi^2$\cr
\hline \hline
1.10 &  8.559     & 4.2114  & 3.7530               & 0.2319     & 9.47               & 2.58    & 1.627\cr
1.12 &  7.766     & 4.2151  & 3.7566               & 0.2504     & 9.51               & 2.59    & 1.670\cr
1.14 &  6.734     & 4.2179  & 3.7607               & 0.2717     & 9.57               & 2.58    & 1.711\cr
1.16 &  6.227     & 4.2197  & 3.7650               & 0.2946     & 9.63               & 2.58    & 1.762\cr
1.18 &  5.802     & 4.2250  & 3.7590               & 0.2727     & 9.65               & 2.60    & 1.849\cr
\hline
\end{tabular}
\end{flushleft}
\end{table*}

They all lie at the beginning of the subgiant branch. For \yg, for both values of the metallicity, the models computed along the evolutionay track have a convective core during the main sequence phase which disappears at the present stage, leaving a helium core with sharp edges. Models with a solar helium abundance \ysol~do not develop a convective core during their evolution, they have a helium core with sharp boundaries.

For a fixed value of [Fe/H], the age of the best model decreases when the helium abundance increases (compare respectively Tables~\ref{tab2} and \ref{tab4} and Tables~\ref{tab3} and \ref{tab5}). This is due to the difference in the evolutionary time scales according to the value of Y. The main sequence time scale increases for increasing helium abundance. As can be seen in Fig.~\ref{fig1}, for a given value of [Fe/H] the effect of a smaller helium abundance is to move the models to lower effective temperatures. For a given value of Y, the models are cooler for a higher metallicity.

\begin{figure*}[btp!]
\begin{center}
\includegraphics[angle=0,totalheight=6.8cm,width=10cm]{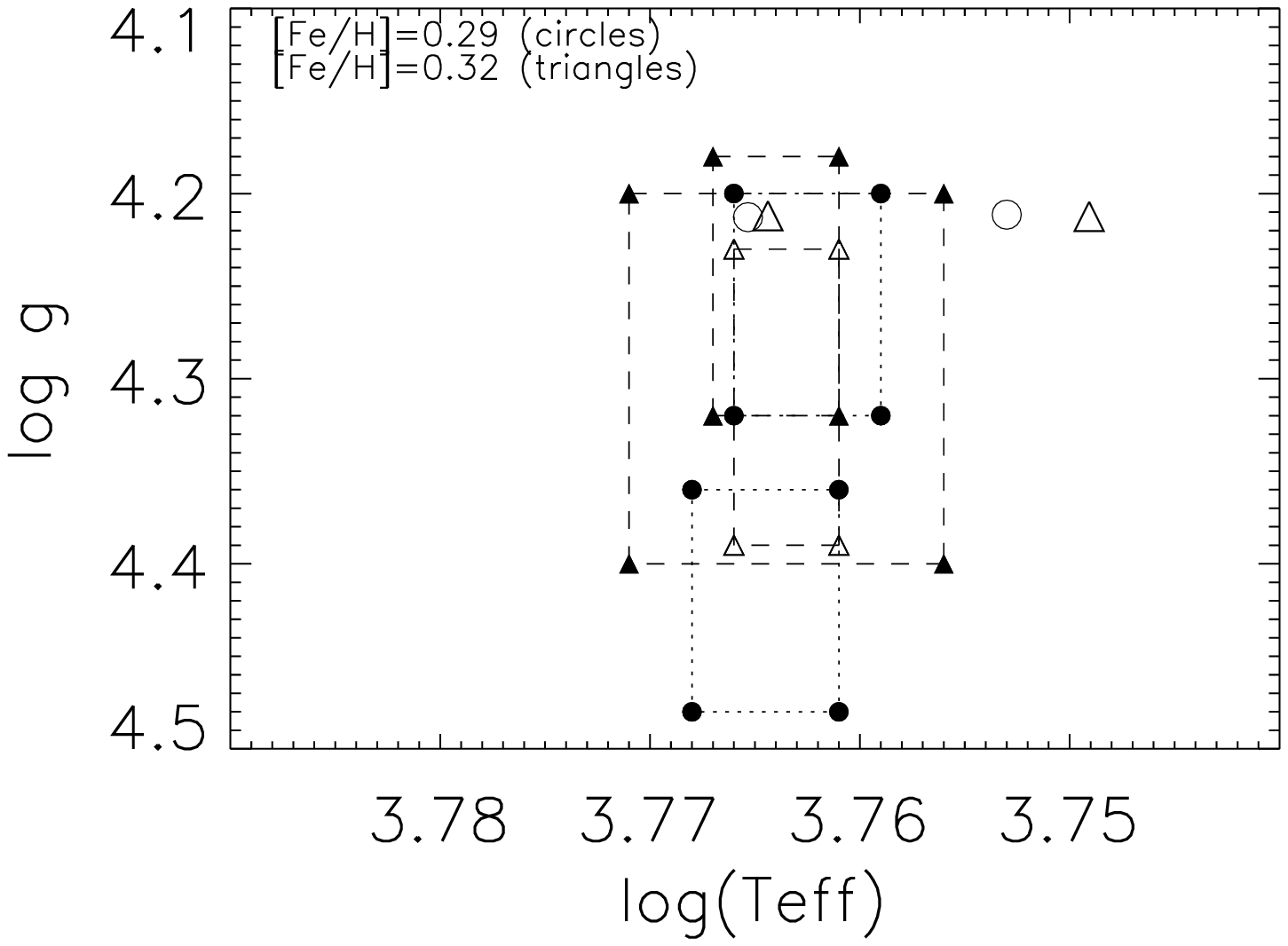}
\caption{Representation of the five spectroscopic error boxes presented in Table~\ref{tab1} for the two values of metallicity: [Fe/H]=0.29 (filled circles): Laws et al. \cite{laws03}, Fischer \& Valenti \cite{fischer05}; and [Fe/H]=0.32 (filled triangles): Bensby et al. \cite{bensby03}, Santos et al. \cite{santos04a} and \cite{santos04b}. The open symbols represent the four best models described in Table~\ref{tab2}. They correspond to the two values of metallicities given above and to two different values of helium abundances: Y$_G$ (Y=0.301) and \ysol~(Y=0.271). For models with the same [Fe/H], the one with a high Y is on the left, and the one with the low Y is on the right.}
\label{fig5}
\end{center}
\end{figure*}

We plotted these four models in a log~$g$~-~log~\teff~plane, together with the spectroscopic error boxes (Fig.~\ref{fig5}). We clearly see that models with the lower helium abundance \ysol~(models~2 and~4) lie outside the spectroscopic error boxes. Their effective temperatures are too low compared to the values derived by the spectroscopists. On the other hand, the models computed with a high helium content (models 1 and 3) lie right inside the error boxes: their external parameters correspond to those derived by spectroscopy. These are also the models with the lowest $\chi^2$ (see Tables~\ref{tab2} to \ref{tab5}). The luminosities of all these models agree with that derived from the Hipparcos parallax, namely log~(L/\lsol)~$=0.25\pm0.03$ (Sect. 2.2). The echelle diagram for the best model 3 is also presented in Fig.~\ref{fig4} in more detail, with all the observed and modelled oscillation modes given separately.

These models differ from the overmetallic model proposed by Bazot et al. (\cite{bazot05}), which corresponded to a more massive star (1.18~\msol) on the main sequence. One of the reasons for this difference is related to the fact that in Bazot et al. (\cite{bazot05}) the luminosity was the basic parameter used for comparisons, and that its value has been modified. The present analysis is more precise and consistent in the comparison with the observations.

Another reason for the different results obtained here is related to the helium content of the star. We find that only models with a high helium value (Y=0.30) correctly reproduce all the observations. The helium abundance, which was kept as a free parameter by Bazot et al. (\cite{bazot05}), is now well constrained. This determination of Y represents an important improvement.

Interestingly enough the case of $\mu$~Arae is different from that of $\iota$~Hor, discussed by Vauclair et al. (\cite{vauclair08}). The asteroseismic study of $\iota$~Hor showed that its helium abundance was low (Y=0.255), comtrary to that of $\mu$~Arae. These two different results have interesting implications for the details of the nuclear processes which lead to the observed overmetallicity. This will be discussed in the conclusion.

\subsection{Constraints on core overshooting}

\subsubsection{Models with \al=0.20 }

The observational echelle diagram presents characteristic variations in the small separations. We were particularly interested in the fact that around $\nu=2$~mHz the lines $\ell=0$ and $\ell=2$ come close to each other, so that the small separation is smaller than average. In a previous paper (Soriano \& Vauclair \cite{soriano08}) we noticed that the small separations between $\ell=0$ and $\ell=2$ may vanish and become negative in stars at the end of the main sequence phase due to their helium cores. In this case there is a crossing point in the echelle diagram, with an inversion between $\ell=0$ and $\ell=2$ modes at high frequencies. 

Here we wanted to check whether adding an overshooting layer at the edge of the convective core could lead to such an effect and reproduce the observed echelle diagram in a different way.
We tried to find a model that could fit the observational echelle diagram and present a crossing point between the lines $\ell=0$~-~$\ell=2$ at a frequency close to 2~mHz. 

\begin{figure*}[btp!]
\begin{center}
\includegraphics[angle=0,totalheight=4.8cm,width=7.0cm]{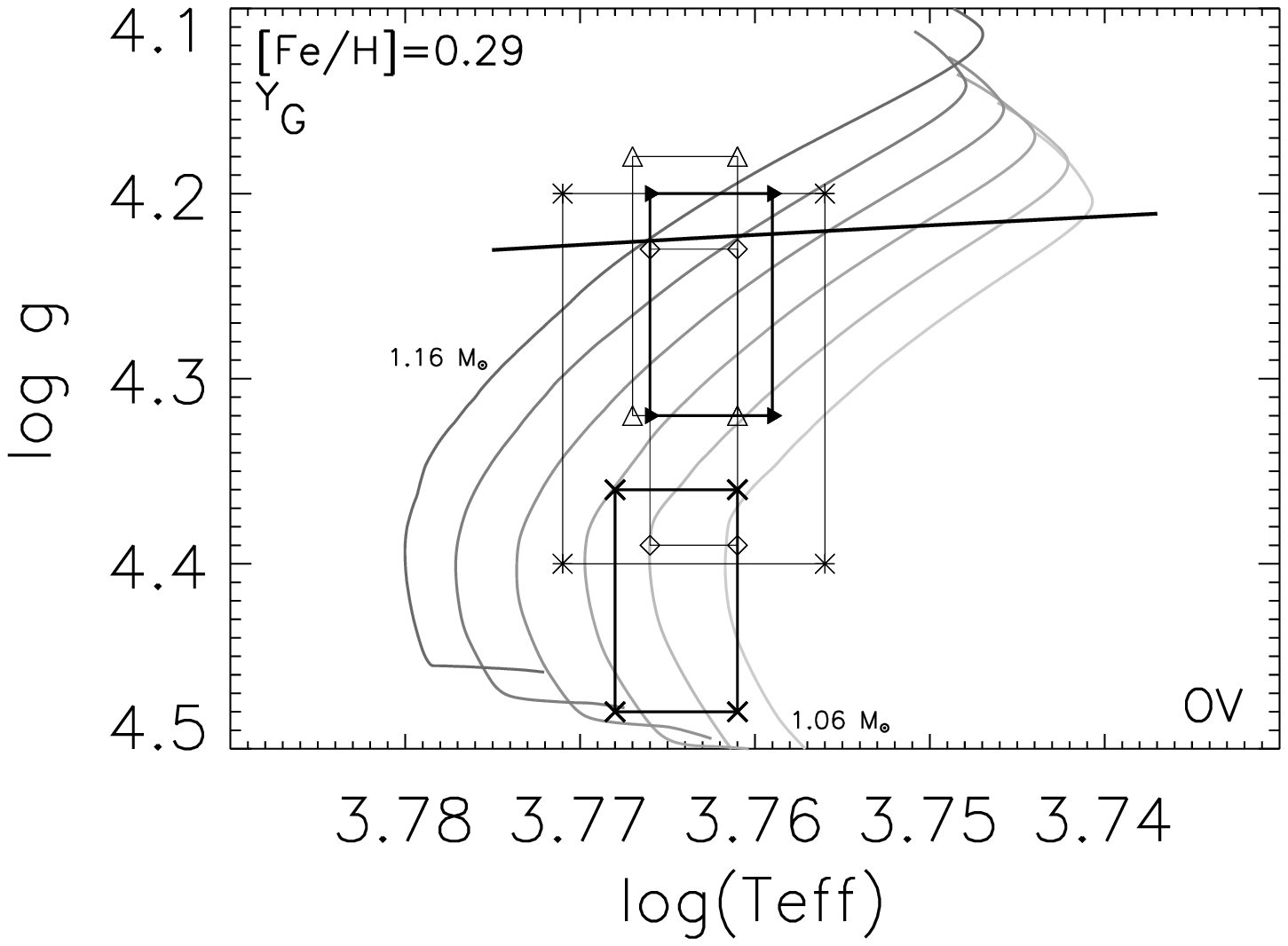}\includegraphics[angle=0,totalheight=4.8cm,width=7.0cm]{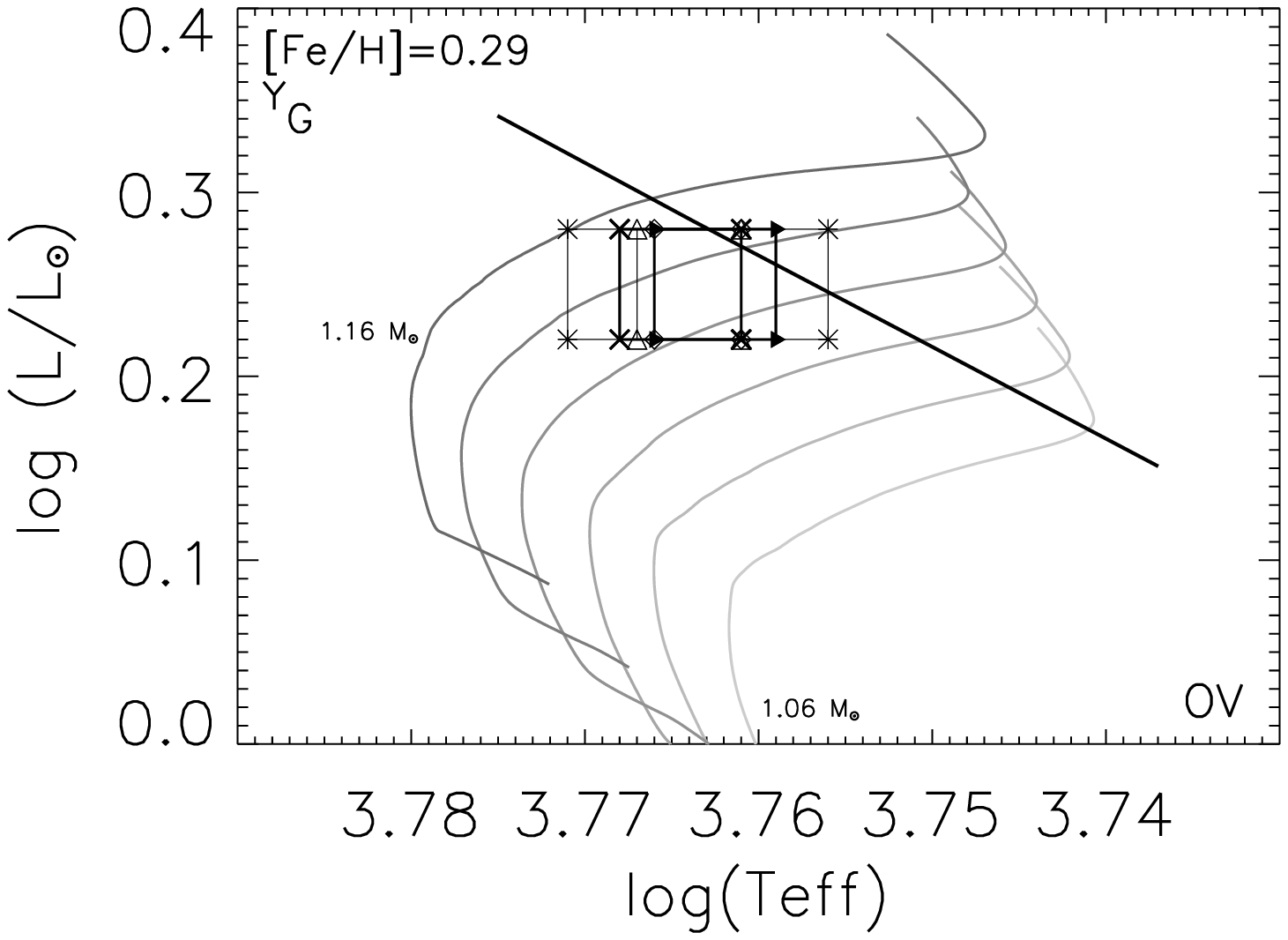}
\includegraphics[angle=0,totalheight=4.8cm,width=7.0cm]{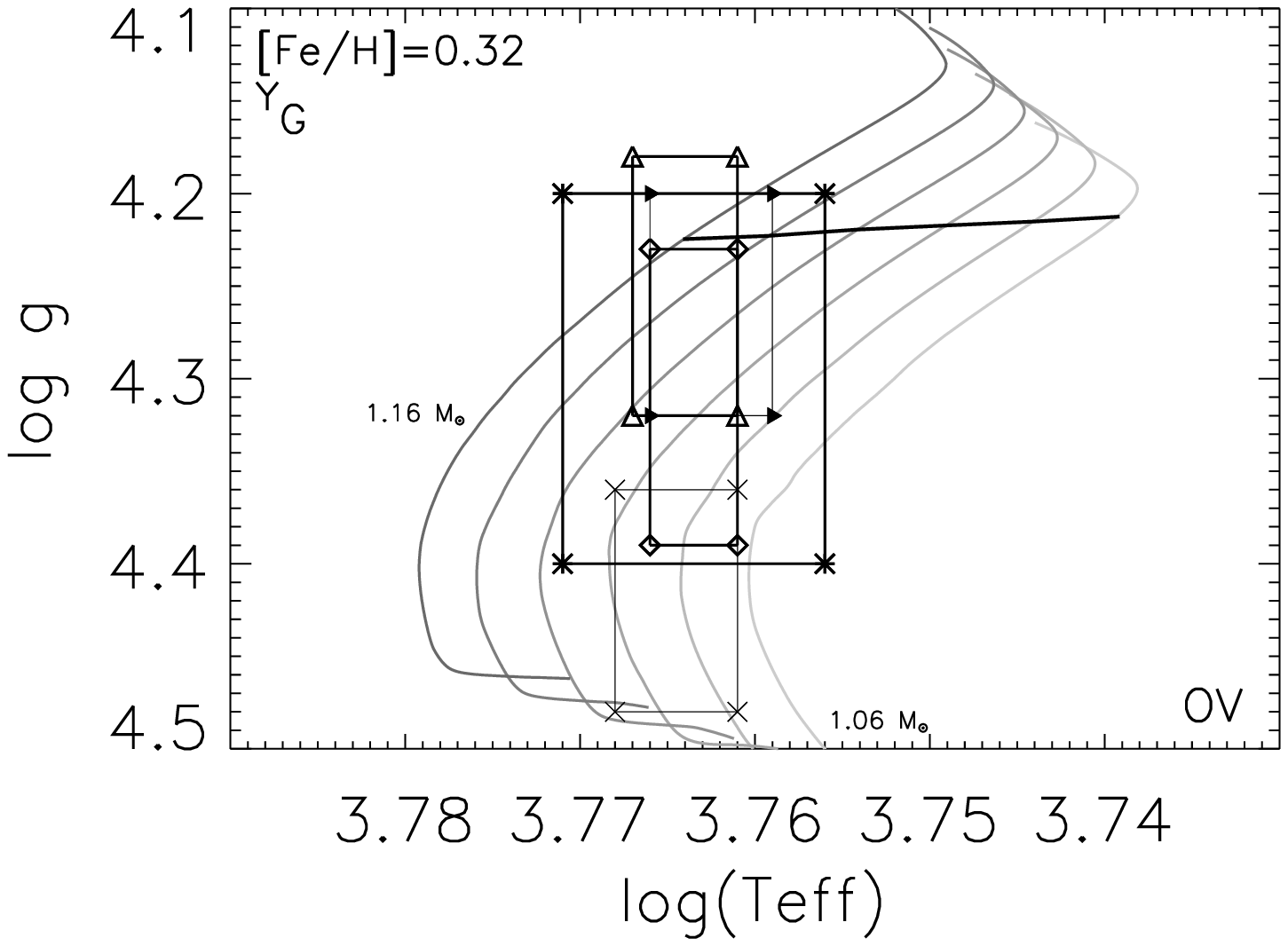}\includegraphics[angle=0,totalheight=4.8cm,width=7.0cm]{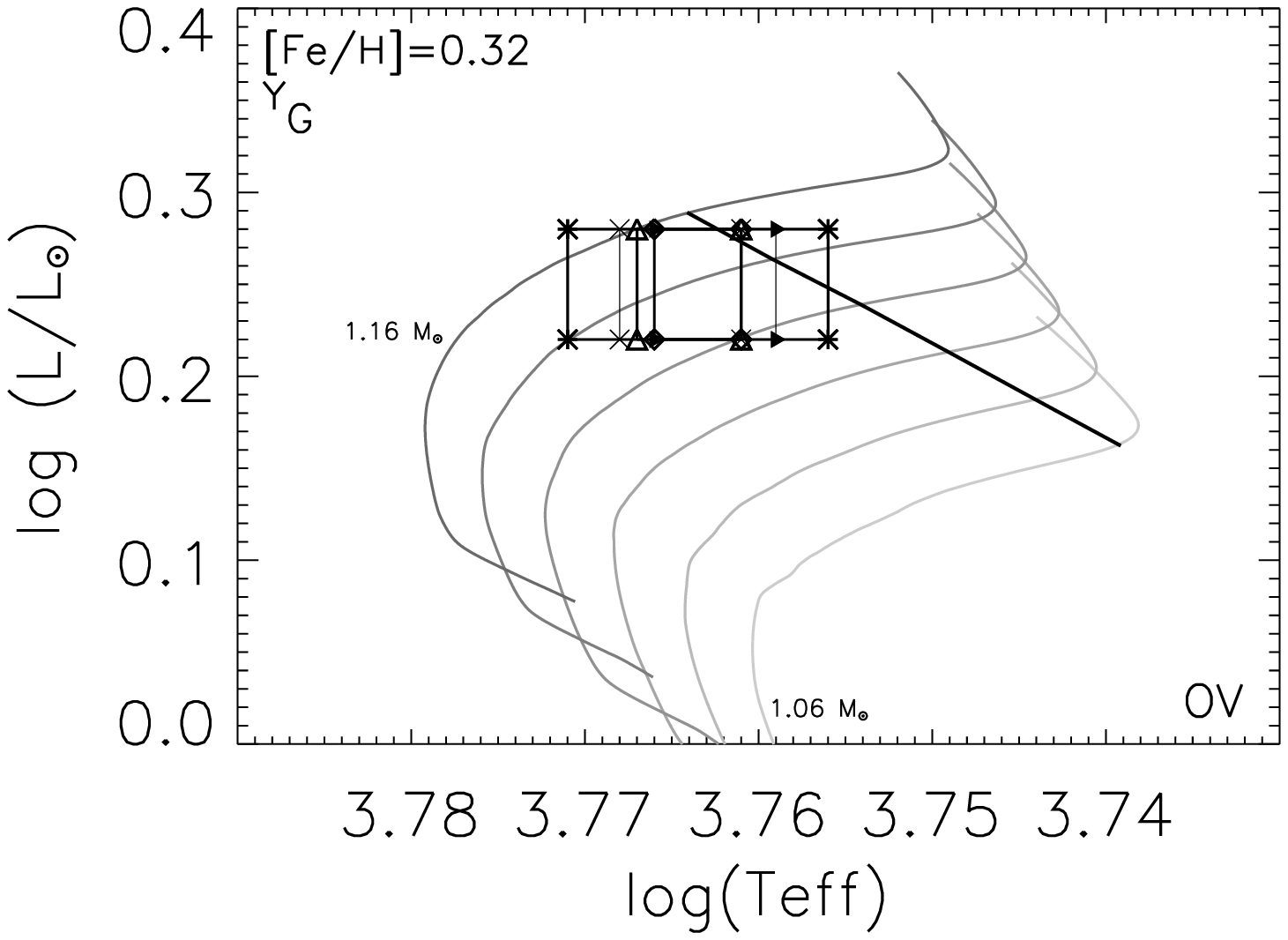}
\includegraphics[angle=0,totalheight=4.8cm,width=7.0cm]{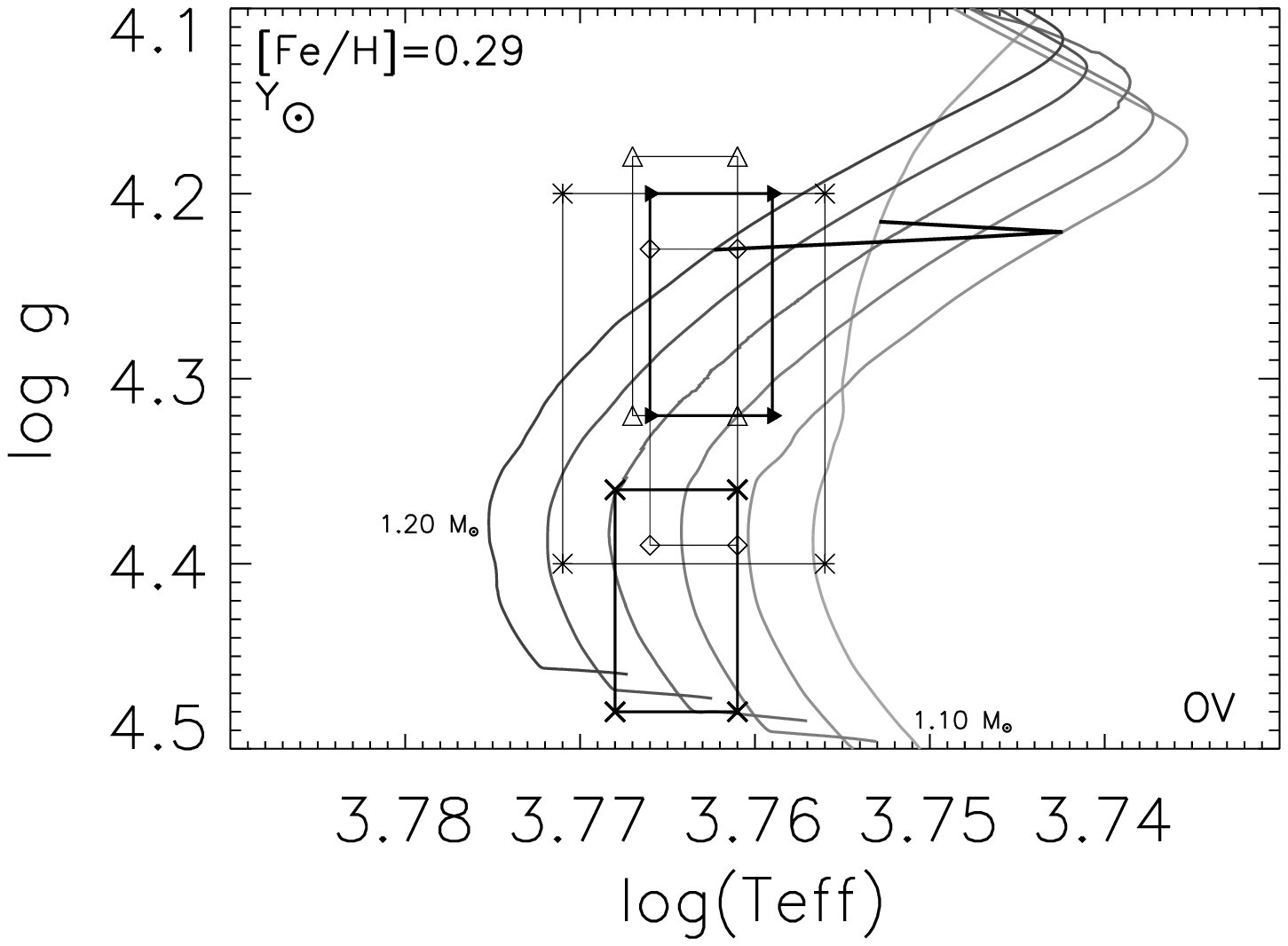}\includegraphics[angle=0,totalheight=4.8cm,width=7.0cm]{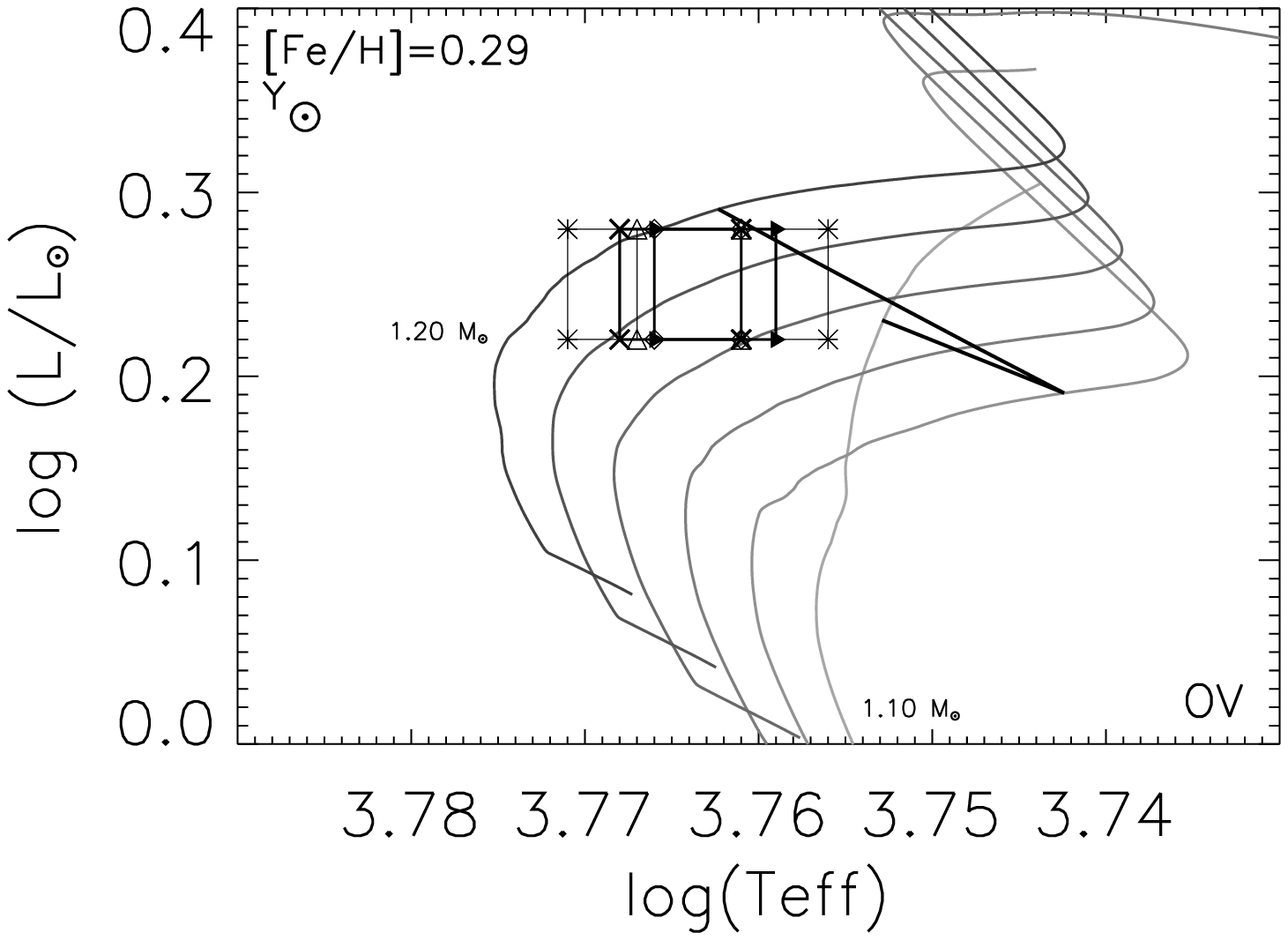}
\includegraphics[angle=0,totalheight=4.8cm,width=7.0cm]{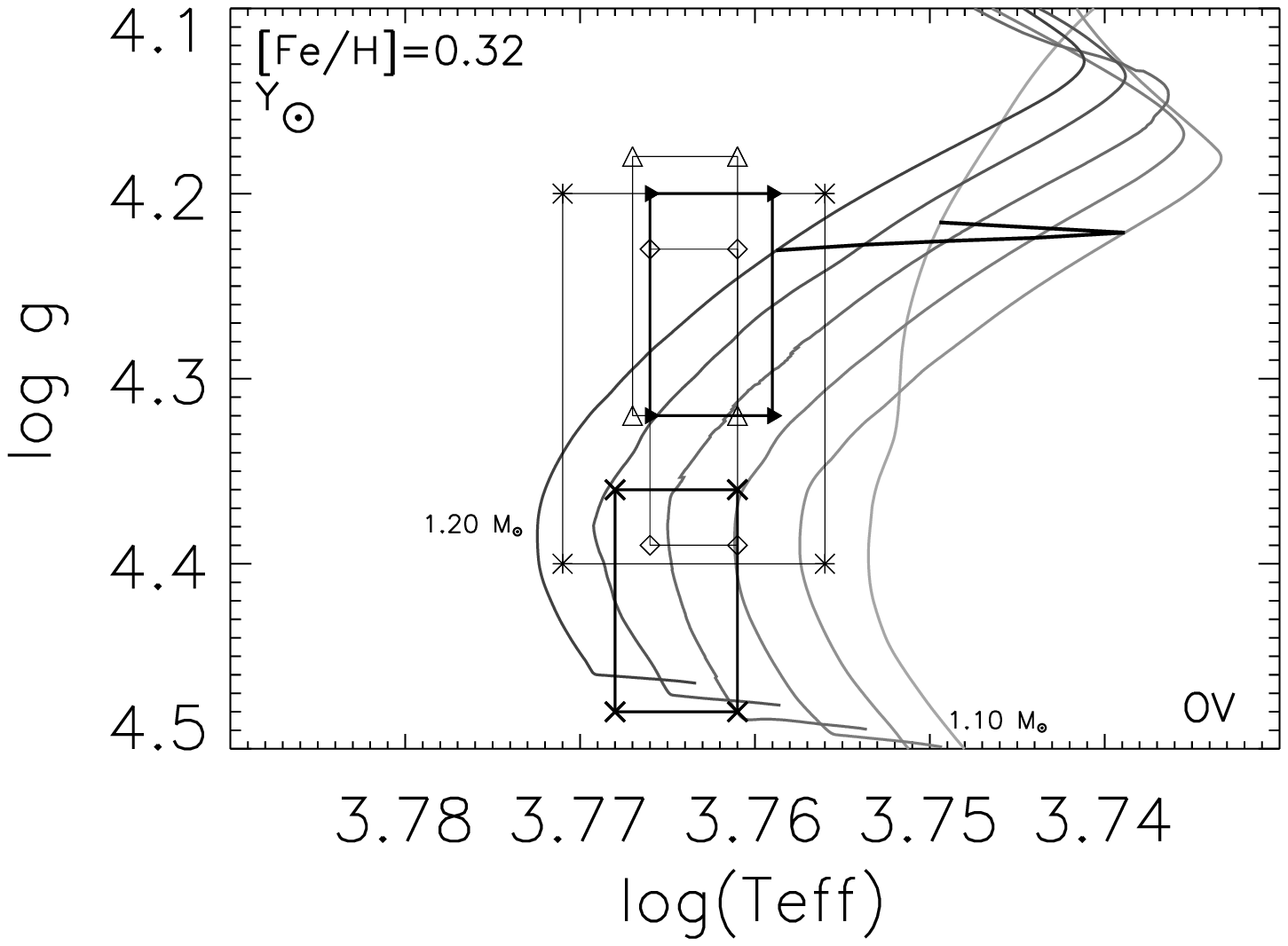}\includegraphics[angle=0,totalheight=4.8cm,width=7.0cm]{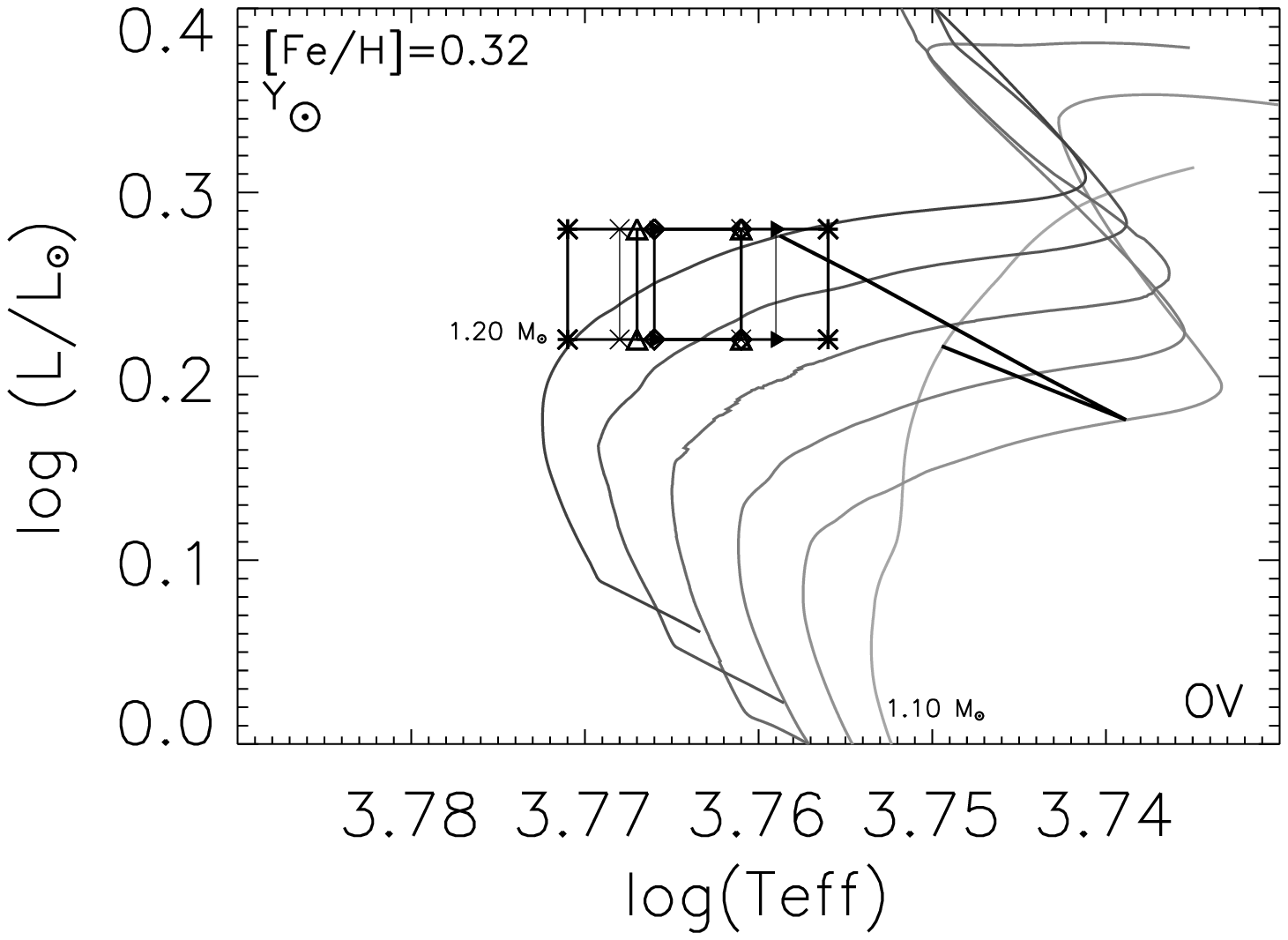}
\caption{Evolutionary tracks in the log~$g$~-~log~\teff~(left panels) and log~(L/\lsol)~-~log~\teff~planes (right panels) for the two values of metallicities found in the literature: [Fe/H]=0.29 and [Fe/H]=0.32, with overshooting at the edge of the core (\al=0.20). The represented error boxes and masses are the same as in Fig.~\ref{fig1}.}
\label{fig6}
\end{center}
\end{figure*}

We computed evolutionary tracks with an overshooting parameter \al~=~0.20 for the two values of metallicity: [Fe/H]=0.29 and 0.32, the two values of helium abundance: \yg~and \ysol, and for masses ranging from 1.06 to 1.20~\msol~(Fig.~\ref{fig6}). As for the cases without overshooting, we picked up the models with a large separation of 90 $\mu$Hz and draw curves of iso large separations. 

In all these models, a convective core develops, even for masses smaller than 1.10~\msol. Models with masses ranging from 1.06 to 1.14~\msol~lie at the end of the main sequence. They have a high central helium abundance Y$_{C}$, between 0.8 and 0.9. Most of these models present a crossing point between the $\ell=0$~and~$\ell=2$ curves in the observed frequency range. However, as can be seen on the example shown in Fig.~\ref{fig8} (upper-left panel), the computed lines do not reproduce the observed ones.

Models with masses higher than 1.14~\msol~correspond to younger stars, still on the main sequence. Their convective core is well developped, but their helium-content Y$_C$ is not high enough to induce a discontinuity in the sound speed profile and so to obtain negative small separations in the considered range of frequencies. There is no crossing point in their echelle diagram. These models do not fit the observational echelle diagram either, as can be seen in the example in Fig.~\ref{fig8} (upper-right panel).

These results show that the observed frequencies in $\mu$ Arae cannot be interpreted in terms of a crossing point between the $\ell=0$~and~$\ell=2$ curves. This is consistent with the interpretation of the doublets observed in the $\ell=2$ curve in terms of rotational splitting (Bouchy et al. \cite{bouchy05}). In this framework, the upper part of this curve could not be misinterpreted as $\ell=0$ frequencies.

As overshooting with \al~=~0.20 is excluded, we decided to test smaller overshooting parameters to obtain a strong constraint on the possibility of overshooting in this star.

\subsubsection{Constraints on the overshoot parameter}

\begin{figure*}[btp!]
\begin{center}
\includegraphics[angle=0,totalheight=6.8cm,width=10cm]{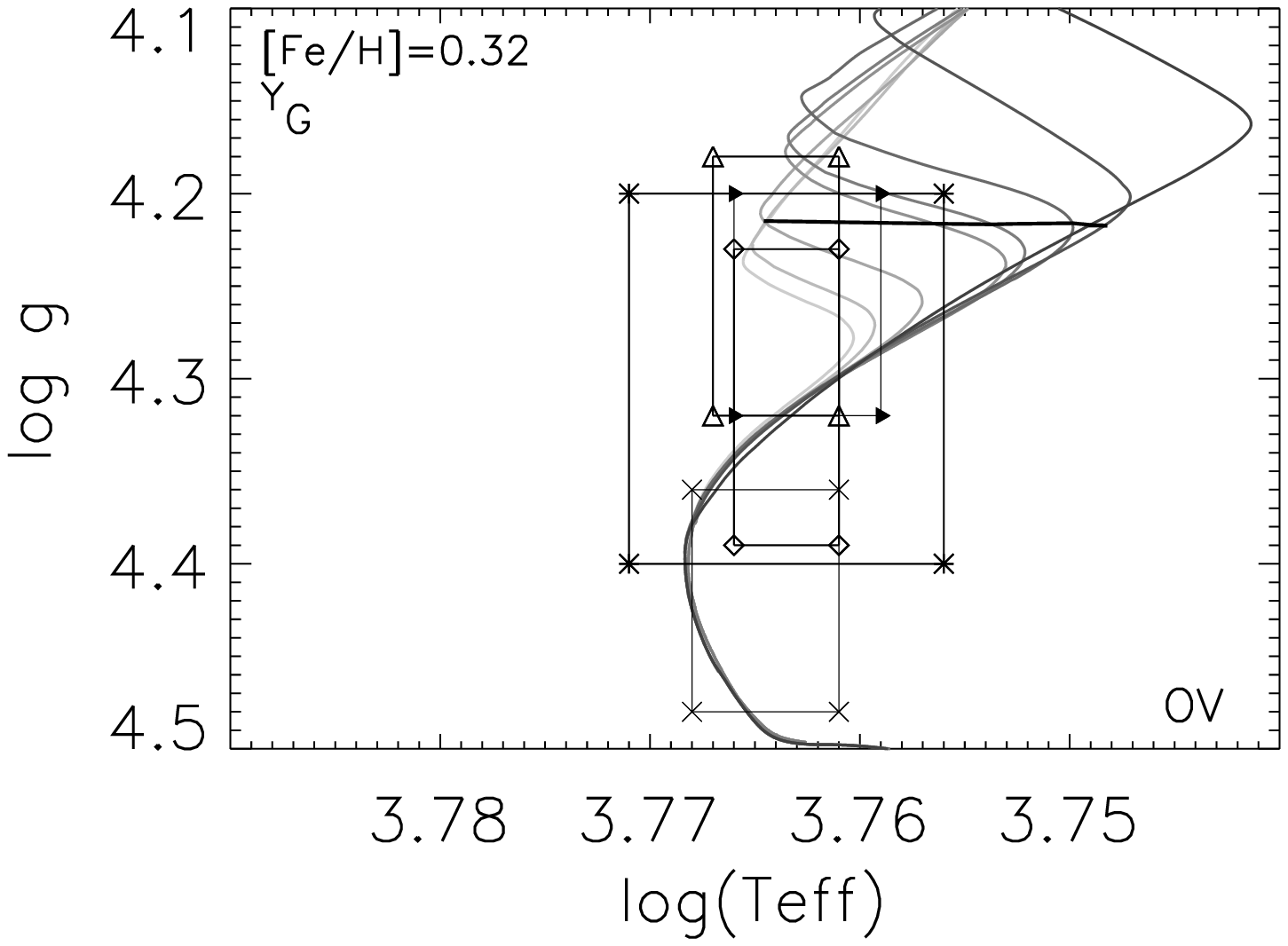}
\caption{Evolutionary tracks for models of 1.10 \msol, with \yg, [Fe/H]=0.32, and a variable overshooting parameter \al. The tracks are for \al= 0, 0.001, 0.002, 0.005, 0.01, 0.04, 0.10, and 0.20. The horizontal thick line represents the iso-$\Delta\nu$ 90$\mu$Hz line. }
\label{fig7}
\end{center}
\end{figure*}

\begin{figure*}
\begin{center}
\includegraphics[angle=0,totalheight=5.5cm,width=8cm]{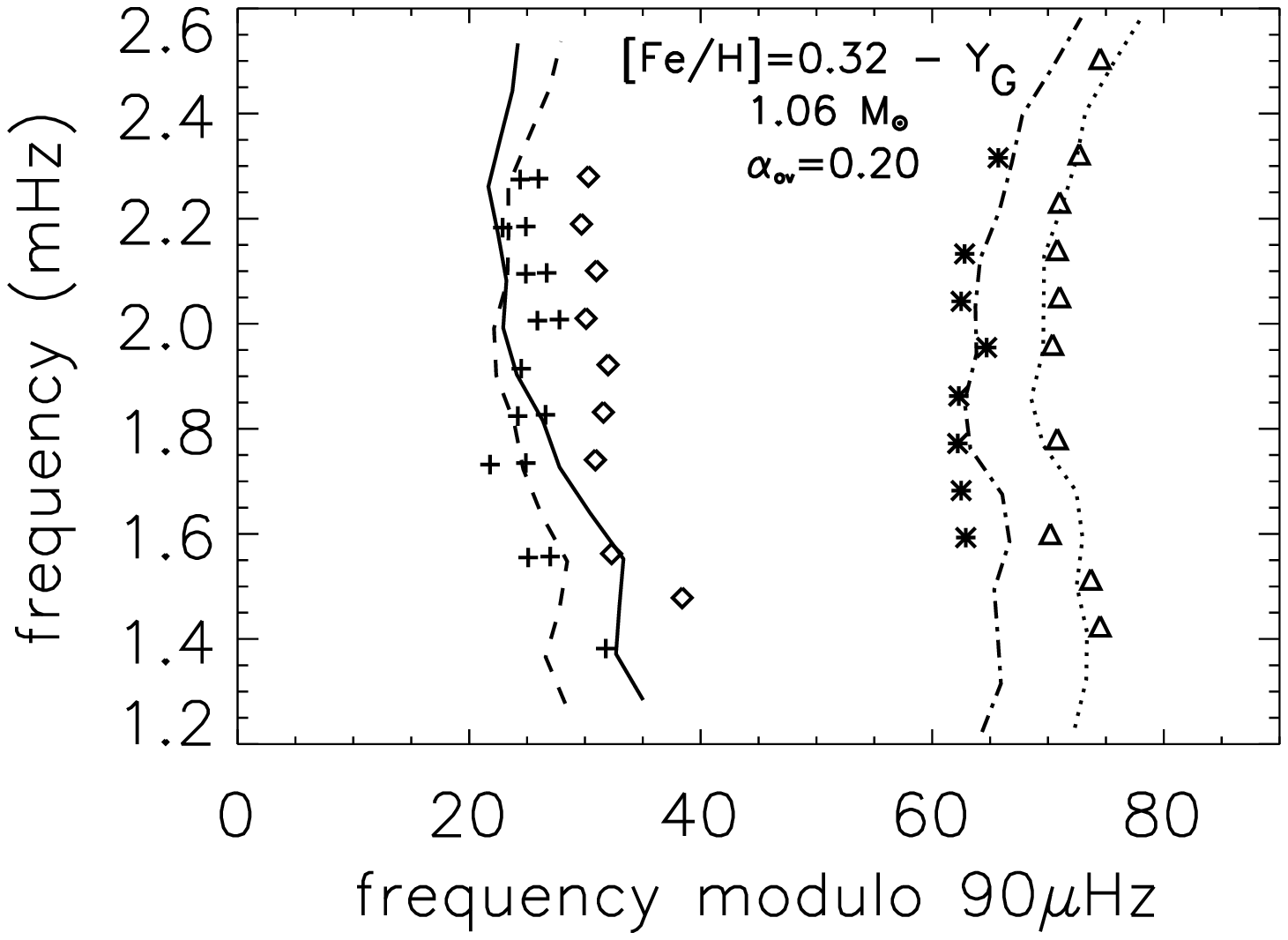}\includegraphics[angle=0,totalheight=5.5cm,width=8cm]{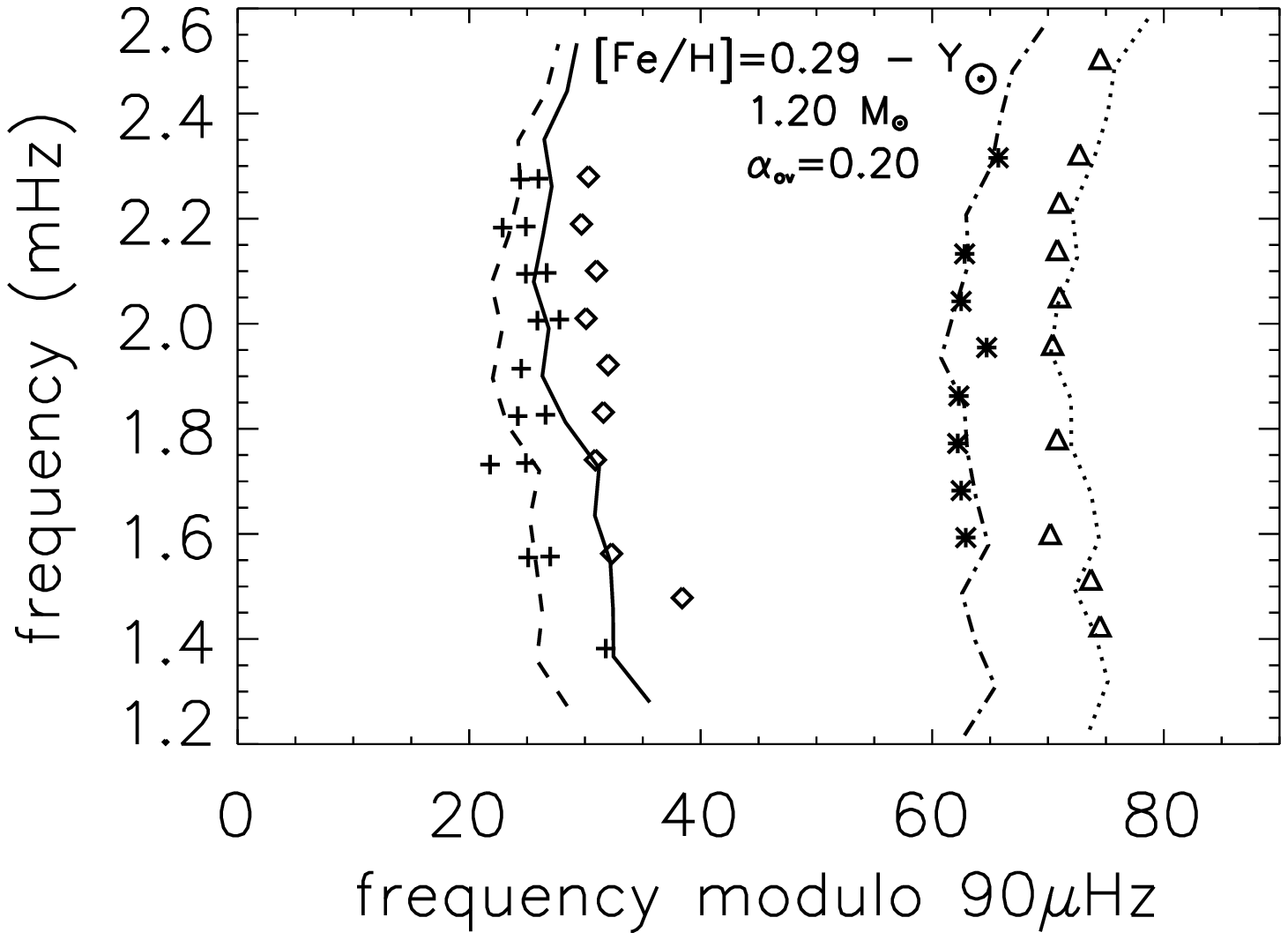}
\includegraphics[angle=0,totalheight=5.5cm,width=8cm]{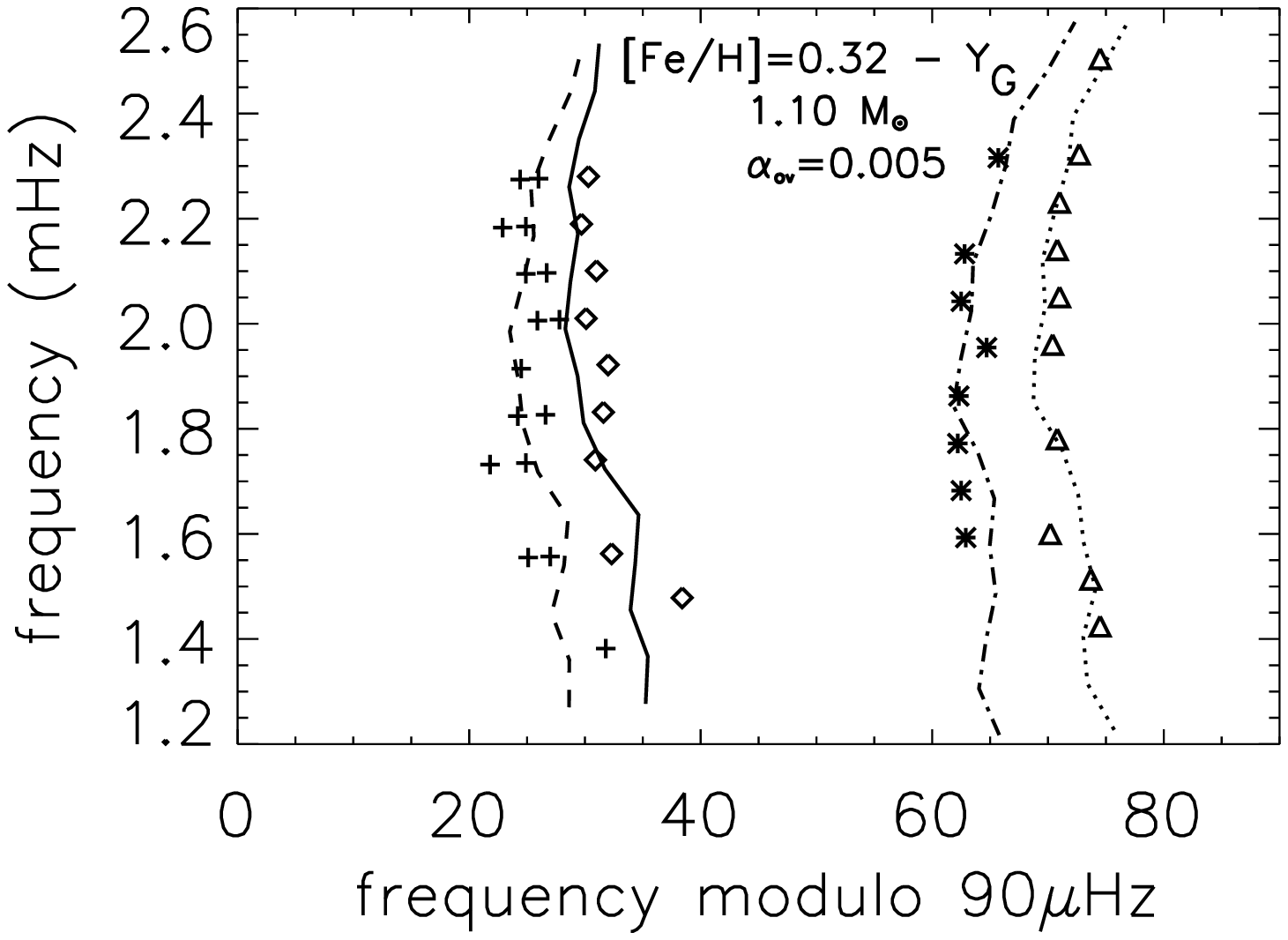}\includegraphics[angle=0,totalheight=5.5cm,width=8cm]{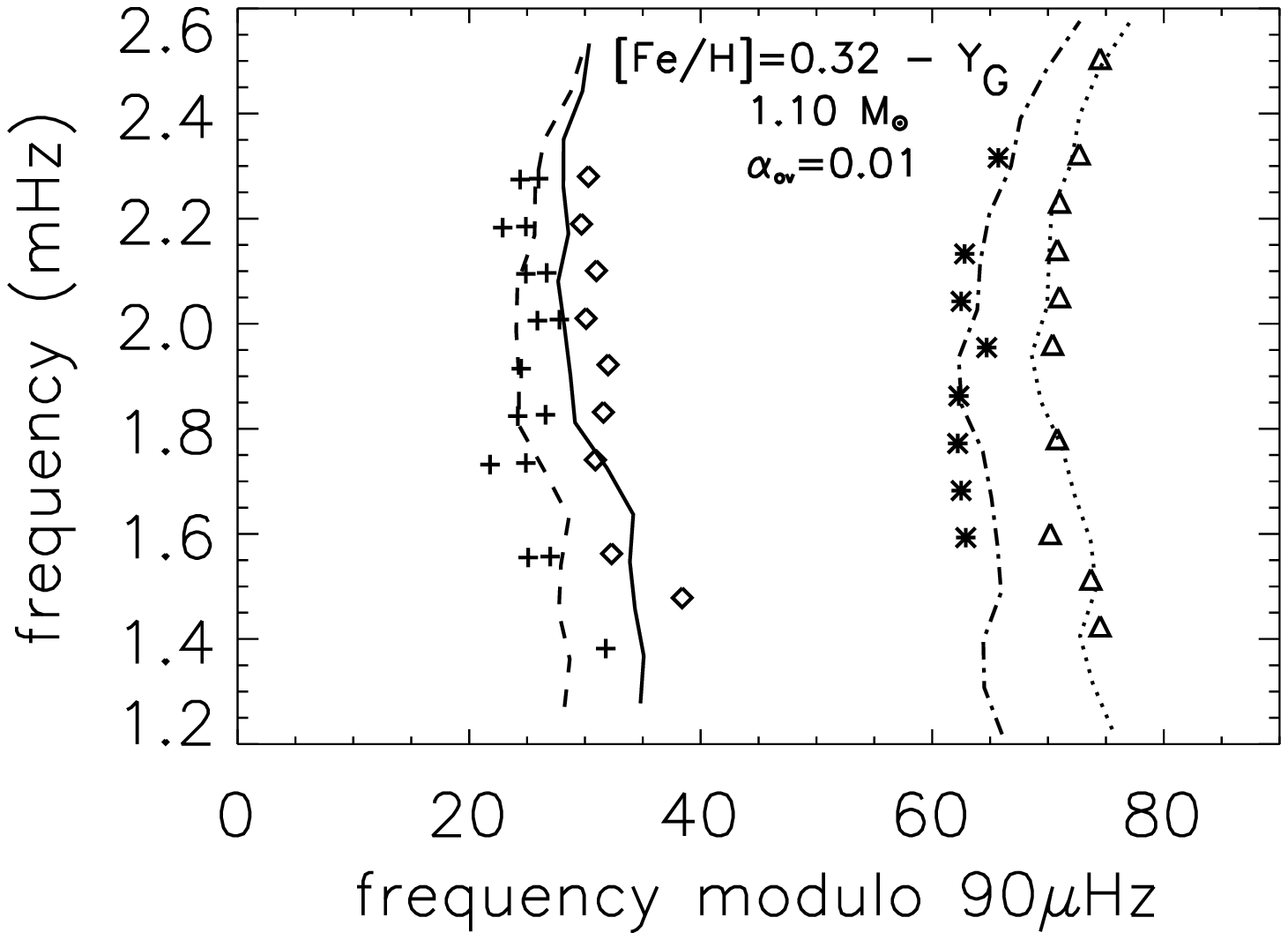}
\includegraphics[angle=0,totalheight=5.5cm,width=8cm]{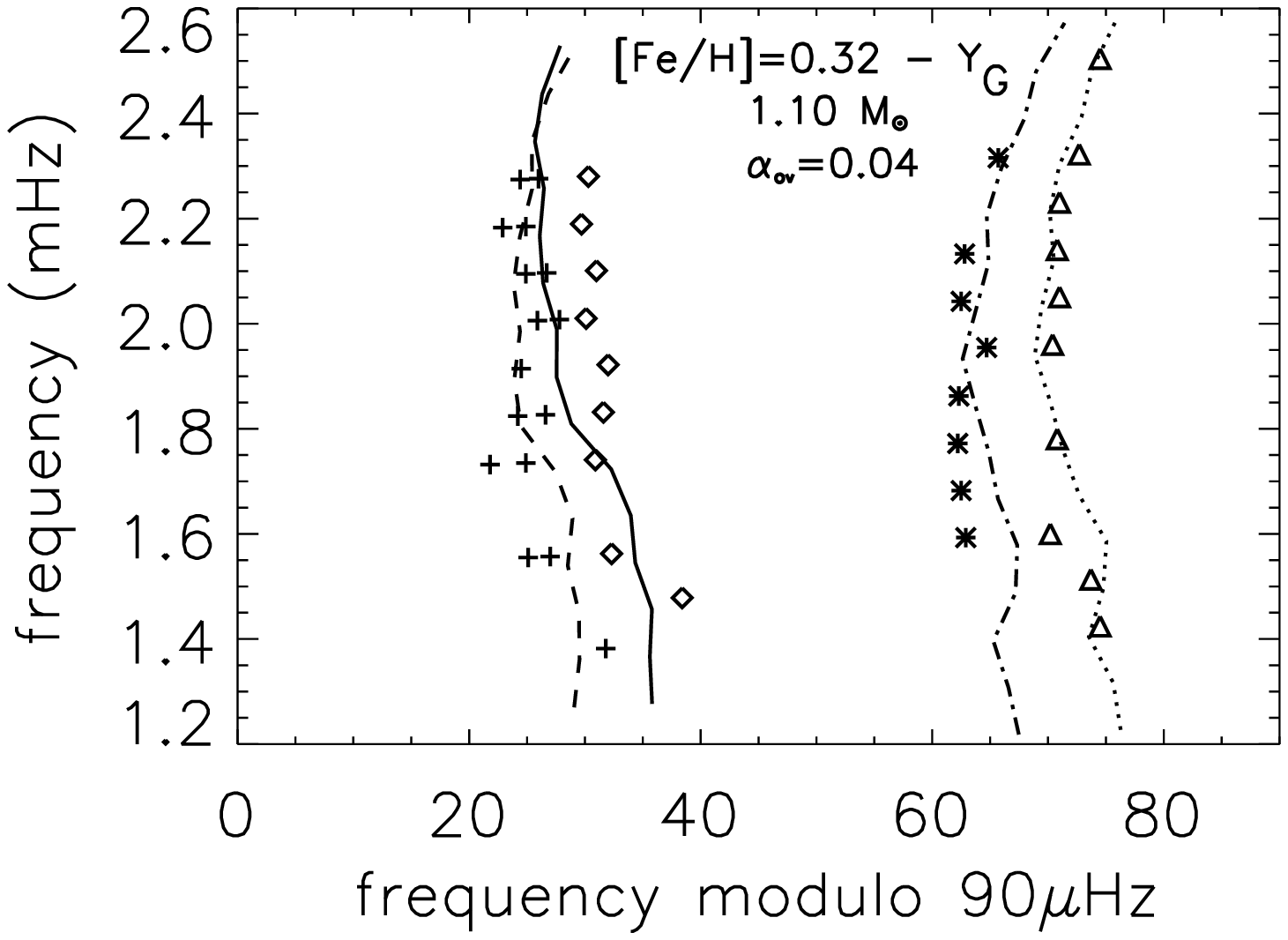}\includegraphics[angle=0,totalheight=5.5cm,width=8cm]{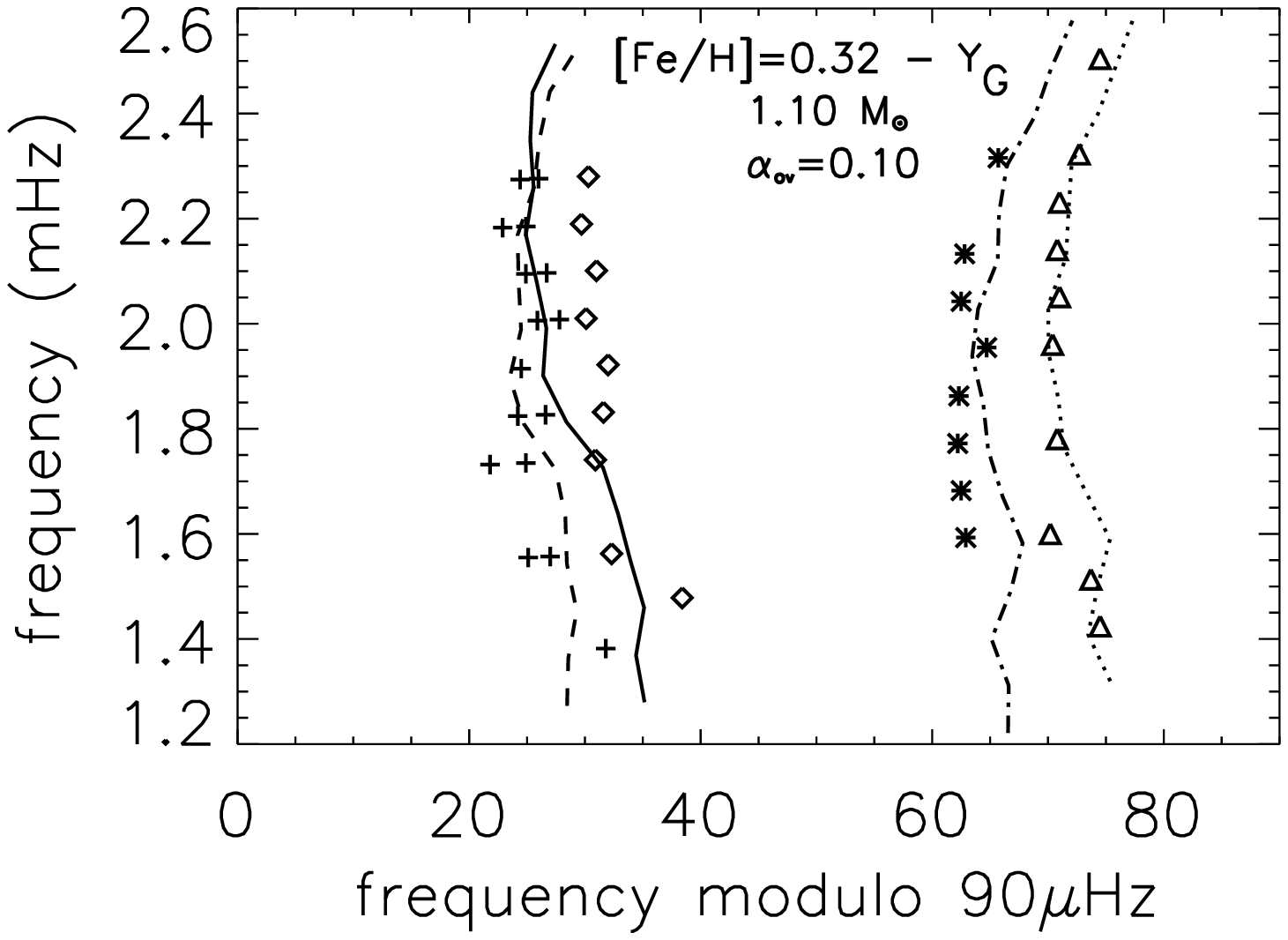}
\caption{Echelle diagram for overmetallic models with overshooting: 1.06~\msol, [Fe/H]=0.32, \yg, and~\al=0.20 (upper-left panel); 1.20~\msol, [Fe/H]=0.29, \ysol, and~\al=0.20 (upper-right panel); 1.10~\msol, [Fe/H]=0.32, \yg, and \al=0.005 (middle-left panel), \al=0.01 (middle-right panel), \al=0.04 (lower-left panel), and \al=0.10 (lower-right panel) . The symbols are the same as in Fig.~\ref{fig3}.}
\label{fig8}
\end{center}
\end{figure*}

We computed new evolutionary tracks for models with 1.10~\msol, [Fe/H]=0.32, and~\yg, gradually increasing the overshoot parameter: \al~=~0.001, 0.002, 0.005, 0.01, 0.04, 0.10 and 0.20. The results obtained in the log~$g$~-~log~\teff~plane are displayed in Fig~\ref{fig7}. Here again we picked up the models with a mean large separation of 90~$\mu$Hz (iso-$\Delta\nu$ line on the tracks). We plotted their echelle diagrams and performed $\chi^2$ tests. The internal characteristics of four of these models are presented in Table~\ref{tab6} and their echelle diagrams are displayed in Fig.~\ref{fig8} (middle and lower panels). Models with \al~=~0.001 and 0.002 are undistinguishable from models without overshooting.

\begin{table*}
\caption{Characteristics of the overmetallic models of 1.10 \msol, [Fe/H]=0.32, for several values of the oovershooting parameter.}
\label{tab6}
\begin{flushleft}
\begin{tabular}{ccccccc} \hline
\hline
\al  & Age (Gyr) & log~\teff & log (L/\lsol) & log $g$ & Y$_{C}$ & r$_{cc}$/R$_{\star}$ \cr
\hline \hline
0.005  & 6.599   & 3.7565    & 0.2484        & 4.2162  & 0.9482  & 0.050              \cr
0.010  & 6.674   & 3.7537    & 0.2323        & 4.2167  & 0.9419  & 0.055              \cr
0.040  & 6.838   & 3.7499    & 0.2177        & 4.2160  & 0.9243  & 0.061              \cr
0.100  & 6.912   & 3.7482    & 0.2096        & 4.2173  & 0.8826  & 0.066              \cr
\hline
\end{tabular}
\end{flushleft}
\end{table*}

The more overshooting is added at the edge of the stellar core, the more the evolutionary time scales are increased, as can be seen in Fig.~\ref{fig7}. The developpement of the convective core is increased during a longer main sequence phase. The models that have a mean large separation of 90~$\mu$Hz are not in the same evolutionary stage, depending on the value of \al, and they do not have the same internal structure. Models with \al $<$ 0.002 are at the beginning of the subgiant branch, as are models without overshooting. When \al~increases, with a value between 0.002 and 0.10, the models are in the phase of contraction of the convective core. And finally, for \al $>$ 0.10, the models are at the end of the main sequence. This difference of evolutionary stages explains that the central helium abundance is lower for models with a higher overshooting parameter.

We studied the evolution of the oscillation frequencies when \al~increases. For small values of overshooting (\al~=~0.001 and \al~=~0.002), there is no visible influence on the oscillations frequencies. 
For \al~=~0.005, the reduced $\chi^2$ is a little higher ($\chi^2=1.66$) than for the model without overshooting. We can see on the echelle diagram (Fig.~\ref{fig8}, middle-left panel) that the lines $\ell=0$~-~$\ell=2$ are closer for high frequencies. For \al~=~0.01, these lines cross at $\nu$~=~2.6~mHz, as can be seen in the echelle diagram (Fig.~\ref{fig8}, middle-right panel). The $\chi^2$ has increased: 1.73. 
For higher values of \al~(\al $>$ 0.01), the crossing point appears for lower frequencies decreasing for increasing \al~(Fig.~\ref{fig8}, lower panels). Their $\chi^2$ value is high (2.45 and 3.04, respectively). These models do not fit the observations. We conclude from these computations that the overshooting parameter is small in $\mu$ Arae, less than 0.01.

Let us recall however that in these computations overshooting is simply treated as an extension of the convective core. We have demonstrated that the seismic analysis of stars is able to give precise constraints on their central mixed zone. This does not exclude other kinds of mild macroscopic motions, plumes for instance, provided they do not lead to complete mixing.

\section{Summary and discussion}

This new analysis of the star $\mu$~Arae confirms that seismology is very powerful and can provide precise values of the stellar parameters with the help of spectroscopy to raise the final degeneracy between the best computed models.

The procedure we use may be summarized as follows.
\begin{itemize}
  \item We need at least the identification of $\ell=0$, 1 and 2 observational modes, and hopefully also $\ell=3$ as in the present paper.
  \item Detailed comparisons of observed and computed modes are necessary. Global values of the so-called large and small separations are not precise enough to derive these parameters. The comparisons of detailed frequencies are needed to take into account the fluctuations and modulations induced by the internal structure of the star, in particular by the central mixed zone.
  \item In this framework, we first compare models and observations in the log~$g$~-~log~\teff~plane. We may check that whatever the chemical composition, models with the same large separation have the same value of the $M/R^3$ ratio, as predicted by the asymptotic theory.
  \item We then go further, by computing for various chemical compositions (Y, [Fe/H]) the model which fits the best the observed echelle diagram, using a $\chi^2$ minimization process.
  \item One of the most important results at this stage, which was already pointed out for the case of the star $\iota$ Hor by Vauclair et al (\cite{vauclair08}), is that these best models, obtained for various chemical compositions, all have the same mass, radius, and thus gravity. With this method, mass and radius are obtained with one percent uncertainty. On the other hand, the age of the star still depends on the chemical composition, basically the helium value.
  \item Then we compare the position of these best models with the spectroscopic error boxes in the log~$g$~-~log~\teff~diagram. This leads to the best choice of the chemical composition. In particular, we now constrain the Y value, which represents an important improvement compared to previous studies.
  \item Finally we check that the luminosity of the best of all models is compatible with that derived from the apparent magnitude and the Hipparcos parallax.
\end{itemize}

\begin{table*}
\caption{Parameters of $\mu$~Arae}
\label{tab7}
\begin{flushleft}
\begin{tabular}{cccc} \hline
\hline
M/\msol        &  1.10~$\pm$~0.02        & \teff~(K)      & 5820~$\pm$50       \cr
R/R$_{\odot}$   &  1.36~$\pm$0.06         & [Fe/H]         &  0.32~$\pm$~0.02   \cr
log~$g$        &  4.215$\pm$0.005        & Y              &  0.30~$\pm$~0.01  \cr
L/\lsol        &  1.90~$\pm$~0.10        & Age (Gyr)      &  6.34~$\pm$~0.80   \cr  
\hline
\end{tabular}
\end{flushleft}
\end{table*}

The parameters found for $\mu$~Arae are given in Table~\ref{tab7}. The uncertainties have been tentatively evaluated by allowing a possible $\chi^2$ increase of 0.1 for each set of models and an uncertainty of 0.1 on the mixing length parameter. They take into account that all computations done with various sets of chemical parameters converge on the same model values. One must keep in mind however that these uncertainties do not include systematic effects which would occur if stellar physics was strongly modified (new opacities, new nuclear reaction rates, new equation of state, etc.).

The new model is different from that given in Bazot et al. (\cite{bazot05}). Apart from the constraint on the Y value, the basic reason is that in this previous paper the comparisons were only based on the stellar luminosity, which was misleading due to the Hipparcos parallax, which was later modified. Also the method used at that time was not as precise as the one we now use. 
Note that the scaling of parameters may lead to wrong results for stars in which the seismic modes cannot be precisely identified . This is the case for example for the mass proposed by Kallinger et al. (\cite{kallinger08}) for the star $\mu$~Arae, which is much too large (1.23~\msol).

We also performed an analysis of the size of the mixed core by testing the implications of overshooting on the mode frequencies. We found a strong constraint on the possibility of core overshooting, treated as an extension of convection: the size of this extension must be less than 0.5~\% of the pressure scale height (overshooting parameter). This does not exclude other kinds of mild boundary effects at the edge of the core, provided that they do not lead to strong mixing.

At the present time, we were able to perform this deep seismic analysis on two solar type stars hosting planets, both with a large metallicity (about twice solar): $\iota$~Hor (Vauclair et al \cite{vauclair08}) and $\mu$ Arae (this paper). In both cases, precise stellar parameters could be obtained. We found however an important difference between these two overmetallic stars. 

In $\iota$~Hor, the helium abundance is low, even lower than the solar value, in accordance with the helium value determined for the Hyades stellar cluster. As other observational parameters also coincide, we concluded that $\iota$~Hor is an ejected member of the Hyades. The reason why the helium abundance is so low in these stars while the metallicity is high is still a mystery, although it certainly depends on the mass of the stars that polluted the original nebula.

In $\mu$~Arae, on the other hand, the helium abundance is large, as expected from the usual laws for the chemical evolution of galaxies (Isotov \& Thuan \cite{isotov04}). This star was formed in a nebula which suffered normal pollution from stars with proportional yields of helium and metals.

Seismology can lead to precise values of helium abundances in solar type stars, where helium cannot be directly derived from spectroscopy. This represents a success, quite apart from all other results and constraints, and will be of importance for the study of the chemical evolution of our Galaxy.

\end{document}